\documentclass[twoside,onecolumn,english,aps,showpacs,prx,superscriptaddress]{revtex4-2}
\usepackage[T1]{fontenc}
\usepackage{bm}
\usepackage{amsmath}
\usepackage{amsthm}
\usepackage{amssymb}
\usepackage{graphicx}
\usepackage{esint}
\usepackage{mathrsfs}
\usepackage{booktabs}
\usepackage{lmodern} 
\usepackage{algpseudocode}
\usepackage{algorithm}
\newtheorem{assumption}{Assumption}
\newtheorem{theorem}{Theorem}

\usepackage{multirow}
\usepackage{makecell}
\usepackage{verbatim}
\usepackage[plainpages = false, pdfpagelabels, bookmarksnumbered = true,
                 breaklinks = true,
                 linktocpage,
                 colorlinks = true,
                 linkcolor = blue,
                 urlcolor  = blue,
                 citecolor = blue,
                 anchorcolor = green,
                 hyperindex = true,
                 hyperfigures
                 ]{hyperref} 
\usepackage{dcolumn}



\begin{document}

\preprint{APS/123-QED}

\title{Topology-Aware Block Coordinate Descent for Qubit Frequency Allocation
 of Superconducting Quantum Processors}

\author{Zheng Zhao}
\affiliation{State Key Laboratory of Low Dimensional Quantum Physics, Department of Physics, Tsinghua University, Beijing, 100084, China}
\affiliation{Frontier Science Center for Quantum Information, Beijing 100184, China}
\author{Weifeng Zhuang}
\affiliation{Beijing Academy of Quantum Information Sciences, Beijing 100193, China}
\author{Yanwu Gu}
\affiliation{Beijing Academy of Quantum Information Sciences, Beijing 100193, China}
\author{Peng Qian}
\affiliation{Beijing Academy of Quantum Information Sciences, Beijing 100193, China}
\author{Xiao Xiao}
\affiliation{Beijing Academy of Quantum Information Sciences, Beijing 100193, China}
\author{Dong E. Liu}
\email{Corresponding to: dongeliu@mail.tsinghua.edu.cn}
\affiliation{State Key Laboratory of Low Dimensional Quantum Physics, Department of Physics, Tsinghua University, Beijing, 100084, China}
\affiliation{Frontier Science Center for Quantum Information, Beijing 100184, China}
\affiliation{Beijing Academy of Quantum Information Sciences, Beijing 100193, China}
\affiliation{Hefei National Laboratory, Hefei 230088, China}

\keywords{quantum computation, block coordinate descent, traveling salesman problem}

\begin{abstract}
    Pre-execution calibration is a major bottleneck for operating superconducting quantum processors, and qubit frequency allocation is especially challenging due to crosstalk-coupled objectives.  We establish that the widely-used Snake optimizer is mathematically equivalent to Block Coordinate Descent (BCD), providing a rigorous theoretical foundation for this strategy for qubit frequency allocation. Building on this formalization, we present a topology-aware block ordering obtained by casting order selection as a Sequence-Dependent Traveling Salesman Problem (SD-TSP) and solving it efficiently with a nearest-neighbor heuristic. The SD-TSP cost reflects how a given block choice expands the reduced-circuit footprint required to evaluate the block-local objective, enabling orders that minimize per-epoch evaluation time. Under local crosstalk/bounded-degree assumptions, the method achieves linear complexity in qubit count per epoch, while maintaining comparable optimization performance. We formalize the calibration objective, clarify when reduced experiments are equivalent or approximate to the full objective, and analyze convergence of the resulting inexact BCD with noisy measurements. Simulations based on a physics-motivated error simulator show that the proposed BCD-NNA ordering attains the same optimization accuracy at markedly lower runtime than graph-based heuristics (BFS, DFS) and random orders, while also achieving optimization quality comparable to a genetic-algorithm baseline. This method is robust to noisy objective-function evaluations and tolerant to moderate non-local crosstalk mismatch. These results provide a scalable, implementation-ready workflow for frequency calibration in near-term superconducting processors and, more broadly, for locality-structured calibration tasks in future scalable architectures.

\end{abstract}

\maketitle

\section{Introduction}

Quantum computing is widely regarded as a transformative technology with the potential to solve certain problems that are difficult for classical computers. 
Representative examples include cryptographic applications \cite{Shor1997}, quantum chemistry and quantum materials simulation \cite{Cao2019,Bauer2020,McArdle2020,Tilly2022}, and broader scientific applications in high-energy physics, many-body physics, and drug design \cite{Di2024,Fauseweh2024,Santagati2024}. 
We also note that near-term quantum algorithms and hybrid quantum-classical methods have been extensively studied as possible routes toward practical applications \cite{Cerezo2021,Pan2024}.
Among the leading hardware platforms, superconducting circuits have demonstrated rapid progress toward scalable and high-fidelity quantum processors. 
This progress includes quantum computational advantage experiments on programmable superconducting devices \cite{Google2019,USTC2021,Zuchongzhi2021,Zuchongzhi2022}, 
as well as more recent advances in large-scale gate optimization, below-threshold error correction, and high-fidelity two-qubit control \cite{Google2024,Google2025,Zuchongzhi2025,Yan2025}.
However, present-day quantum processors still operate far from fully fault-tolerant scale and are characterized by significant noise \cite{Preskill2018,Endo2021}. 
In both current pre-fault-tolerant devices and future scalable architectures, developing effective calibration strategies to achieve high-fidelity quantum operations remains critical for advancing practical quantum computing.

However, as the number of qubits in a quantum processor increases, 
the calibration of system parameters becomes a challenging task\cite{Fedorov2019,Kliesch2021}. 
The entanglement in quantum systems and the potential crosstalk between control parameters result in an exponential growth in the search space for optimization,
leading to the difficult in identifying optimal or local optimal within a reasonable time frame. 
Furthermore, the complexity of quantum circuits used for calibration grows exponentially with the number of qubits, further complicating the optimization process. 
Therefore, efficient multi-parameter optimization algorithms are essential to ensure the calibration of large-scale quantum systems within practical constraints.

The calibration and benchmarking of quantum processors rely on a broad set of characterization tools, including randomized benchmarking and its variants \cite{Knill2008,Magesan2012,Helsen2019,Proctor2019,Helsen2022}, 
cross-entropy and cycle-based benchmarking methods \cite{Boixo2018,Erhard2019,Gu2023}, 
and more general discussions of certification and capability assessment \cite{Eisert2020,Salonik2021,Proctor2022,Proctor2025}. 
Within this broader calibration workflow, qubit frequencies play a significant role in determining the fidelity of superconducting quantum circuits \cite{Bal2024,Wang2022,Place2021,Burnett2019}.
Frequency control impacts not only individual qubit performance but also the mitigation of crosstalk between qubits, 
which is critical for achieving high-fidelity quantum gates \cite{Rol2020, Krantz2019,Koch2007}. 
Previous studies have attempted to address frequency allocation problems using artificial intelligence techniques \cite{Liu2025} or optimization methods, 
such as the Snake optimizer \cite{Google2024, Klimov2020}.
These approaches address different layers of the problem. Machine-learning-based parameter-design methods aim to predict or learn favorable parameter configurations from data, whereas the present work focuses on the optimization architecture itself: how to decompose a crosstalk-coupled objective into tractable block-local subproblems and how to choose a low-cost traversal order for those subproblems. The two directions are therefore complementary rather than direct algorithmic substitutes.
The Snake optimizer, while presented as a graph-traversal-based calibration strategy, is fundamentally an instance of the Block Coordinate Descent (BCD) algorithm applied to qubit frequency allocation in superconducting quantum processors. In this work, we make this connection explicit by establishing the mathematical equivalence between Snake and BCD, providing a rigorous theoretical framework enabling the application of classical optimization theory to this problem. This formalization reveals that the block traversal order---treated as a heuristic choice in the original Snake proposal---is a key degree of freedom that can be systematically optimized. 
A broad optimization literature is also relevant to this problem, including zeroth-order and stochastic nonconvex optimization \cite{Ghadimi2013,Arjevani2023,Khaled2020}, 
complexity analyses for oracle-based optimization \cite{Agarwal2012}, 
and inexact or block-structured minimization methods \cite{Razaviyayn2013,Blake2016}. 
However, directly applying these generic methods to large superconducting devices can still be challenging because the local objective evaluations themselves are expensive and constrained by hardware structure, highlighting the need for more topology-aware and scalable strategies.

In this work, we focus on the qubit frequency allocation problem in superconducting quantum processors and propose an enhanced optimization framework that combines Block Coordinate Descent (BCD) with the Nearest Neighbor Algorithm (NNA).
Our approach leverages the topology of quantum chips and crosstalk structure to significantly reduce optimization complexity while maintaining accuracy. 
The contributions of this study are as follows: 
\begin{enumerate}
    \item We establish the mathematical equivalence between the Snake optimizer \cite{Klimov2020,Google2024} and Block Coordinate Descent (BCD), providing a rigorous theoretical foundation for this widely-used strategy for qubit frequency allocation and enabling the application of classical optimization theory to this problem.
    \item We formulate the block ordering problem as a Sequence-Dependent Traveling Salesman Problem (SD-TSP) and solve it efficiently using the Nearest Neighbor Algorithm (NNA). This principled approach yields systematic complexity reduction compared to the graph-based heuristics (BFS, DFS) employed in the original Snake implementation.
    \item We provide a rigorous complexity analysis demonstrating $\mathcal{O}(N)$ scaling under local crosstalk assumptions. We further characterize the benefits of ordering optimization under two complexity models: an empirical model reflecting practical implementations, and the search-space model from Ref.~\cite{Klimov2020} representing theoretical worst-case complexity.
    \item Through extensive numerical simulations, we validate the convergence, efficiency, and robustness of the proposed algorithm under measurement noise and non-local crosstalk conditions.
\end{enumerate}
This study offers a new perspective on calibration strategies for near-term superconducting processors and for larger-scale architectures that retain locality in their control and crosstalk structure,
paving the way for more efficient and scalable approaches to enhance quantum system performance.

The remainder of the paper proceeds as follows. We first cast qubit frequency allocation into a BCD framework that makes the objective, the local surrogate construction, convergence conditions, and complexity measures mathematically precise; this step also clarifies that the Snake strategy is one instance of a broader design space. Within this framework the block-update order emerges as an explicit degree of freedom, which we optimize through an SD-TSP formulation solved by a nearest-neighbor heuristic. Numerical experiments on a physics-motivated error simulator then test whether the resulting structural improvement yields practical gains in optimization quality, algorithmic cost, and robustness.

\section{Results}
Before presenting numerical results we need to fix several ingredients: the optimization framework itself, the locality assumptions that govern reduced-objective construction, and the error simulator together with its performance metrics. These definitions determine both what is being optimized and what cost measure is meaningful. The quantitative validation that follows---covering optimization improvement, robustness, and complexity reduction---should therefore be read against the methodological choices laid out first.

Our methodology for qubit frequency allocation addresses the challenge of exponential complexity by decomposing the global optimization problem. This approach begins by leveraging preliminary characterization data, including the processor's physical topology and a map of its crosstalk interactions, to partition the system's parameters into manageable blocks. For each block, we then define a local objective function evaluated through a reduced-scale calibration experiment. The guiding principle is to construct this experiment using the minimum number of qubits necessary to capture the full impact of the parameters within the block, including all significant crosstalk effects. This ''crosstalk footprint'' is determined from prior knowledge of the system; for example, a local crosstalk model assumes interactions are confined to physically adjacent qubits, whereas a more complex model might account for non-local couplings.

By restricting the calibration experiment to this minimal qubit footprint, we ensure the objective function remains computationally tractable, as its evaluation complexity typically scales exponentially with the number of qubits involved. While recent work \cite{Google2024,Liu2025} has demonstrated the use of machine learning to generate surrogate objective functions from experimental data, our strategy of defining a reduced objective based on a physical crosstalk model is a complementary and vital technique for managing complexity. Even when a global objective function is learned, its evaluation for a local parameter update can be made vastly more efficient by considering only the terms relevant to that block's crosstalk footprint.

This framework of sequential, block-wise optimization is implemented using the Block Coordinate Descent (BCD) algorithm. The overall efficiency of this process is critically dependent on the strategy used to partition parameters into blocks and the sequence in which these blocks are optimized. An effective blocking and ordering strategy, as we will detail, minimizes the computational cost per optimization epoch, enabling the efficient enhancement of quantum circuit fidelity.

\begin{figure*}
    \centering
    \includegraphics[width=1.0\linewidth]{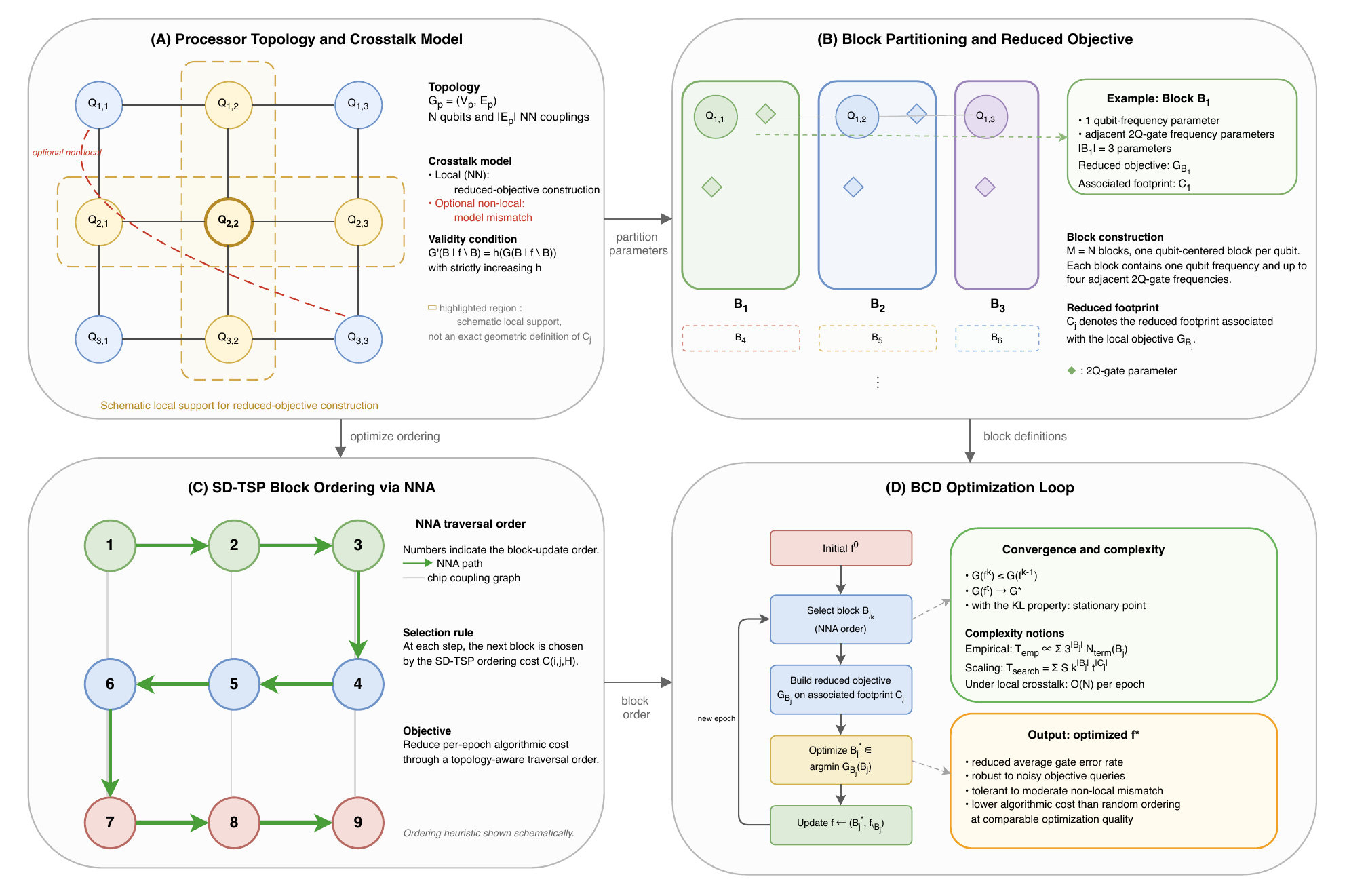}
    
    \caption{
\textbf{Topology-aware BCD-NNA framework for qubit frequency allocation.}
(A) A superconducting processor is represented by a nearest-neighbor chip graph \(G_p=(V_p,E_p)\). The dashed highlighted regions schematically indicate the local support used to construct reduced objectives under a local-crosstalk assumption, while the dashed red link illustrates optional non-local crosstalk that is not included in the reduced model and therefore acts as a source of model mismatch. The condition \(G'(B\,|\,f\backslash B)=h(G(B\,|\,f\backslash B))\) with strictly increasing \(h\) indicates when the reduced local objective preserves the ordering induced by the full objective for block updates. (B) The tunable frequency-related parameters are partitioned into qubit-centered blocks \(B_j\). Each block contains one qubit-frequency parameter together with adjacent two-qubit-gate frequency parameters; for example, block \(B_1\) contains 3 parameters. For each block \(B_j\), the optimizer constructs a reduced local objective \(G_{B_j}\) on an associated reduced footprint \(C_j\). (C) The block-update order is formulated as a sequence-dependent traveling salesman problem and solved using a nearest-neighbor algorithm. The numbered nodes and green arrows illustrate one traversal order chosen to keep successive reduced objectives spatially compact and thereby reduce per-epoch algorithmic cost. (D) Starting from an initial parameter vector \(f^0\), the BCD loop repeatedly selects a block, constructs \(G_{B_j}\), solves the corresponding block-local minimization problem, and updates the global parameter vector. The right-hand panels summarize the convergence and complexity notions used in this work, including monotonic decrease under valid local surrogates, convergence to a stationary point under additional assumptions, the empirical cost proxy \(T_{\mathrm{emp}}\propto \sum_j 3^{|B_j|}N_{\mathrm{term}}(B_j)\), the search-space model \(T_{\mathrm{search}}=\sum_j S\,k^{|B_j|}t^{|C_j|}\), and the resulting \(O(N)\) per-epoch scaling under local crosstalk.
}
    \label{fig:flowchart}
\end{figure*}

\subsection{The role of crosstalk in local objective design}

The central challenge in optimizing a specific block of parameters, denoted as $B$, is that adjustments can have unintended consequences on other parts of the quantum circuit. This phenomenon, known as crosstalk, arises from unwanted physical interactions, such as parasitic flux coupling or residual $ZZ$ interactions between qubits. To manage this complexity, our block coordinate descent approach relies on a local calibration experiment, $G'$, designed to be computationally tractable while accurately capturing the effects of varying the parameters in $B$.

For this local experiment to be valid, its optimization landscape must be consistent with that of the full, global calibration experiment, $G$. Formally, $G'$ preserves the ordering induced by $G$ when all other parameters are held constant. This validity condition is expressed as:
\begin{equation}
    G'(B|f\setminus B)=h(G(B|f\setminus B))
    \label{eq:optimal}
\end{equation}
where the function $h$ is strictly increasing, $B$ represents one block of parameters in chip and $f$ represents the set of all parameters. This condition ensures that a step taken to optimize the local experiment does not contradict the objective of the global optimization.

The validity of the inclusion condition in Eq.~(\ref{eq:optimal}) depends entirely on the construction of the local experiment $G'$. Specifically, $G'$ must incorporate all gates whose performance is affected by the parameters in block $B$. Identifying this set of gates requires an underlying crosstalk model, a set of assumptions about which components in the system interact. Two principal approaches exist for defining this model.

The first approach is to use an exhaustive model, which assumes all-to-all connectivity. This defines an inclusion range for $G'$ that encompasses every qubit in the processor. Such a comprehensive construction guarantees that all crosstalk effects are captured, ensuring the validity of Eq.~(\ref{eq:optimal}) by definition. While theoretically sound, this method is only practical for very small quantum systems, as the experimental and computational complexity scales intractably with the number of qubits. For larger systems, its value is primarily theoretical, confirming that a valid experiment $G'$ always exists.

The second, more practical approach is to employ a heuristic model. This model is typically based on physical priors, such as the chip's known connectivity graph (e.g., nearest-neighbor coupling), and is refined with data from preliminary characterization experiments. This leads to a tractable experiment $G'$ with a limited scope. However, this strategy introduces the critical risk of a model-reality mismatch. If the heuristic model is an incomplete or inaccurate representation of the true physical crosstalk, then the experiment $G'$ will fail to account for all relevant interactions. Consequently, the inclusion condition in Eq.~(\ref{eq:optimal}) may be violated, causing the local optimization to converge towards a solution that is suboptimal for the global system. The iterative process of detecting further crosstalk and expanding the experiment's scope corresponds to an attempt to refine this heuristic model until the validity condition is met.

When employing a heuristic model, the iterative process of refining it to account for physical crosstalk can lead to intermediate stages where a model-reality mismatch persists. In these scenarios, the inclusion condition in Eq.~(\ref{eq:optimal}) is only approximately satisfied, indicating that the local optimization may converge to a suboptimal solution for the global system. The optimization algorithm can then proceed via one of two adaptive strategies:
\begin{enumerate}
    \item Continue the iterative refinement of the heuristic model. This involves detecting further crosstalk effects and expanding the experiment's scope ($G'$) until the inclusion condition in Eq.~(\ref{eq:optimal}) is rigorously met.
    \item Halt the refinement process and proceed with the optimization using the current, tractable heuristic model. This approach accepts the existing model-reality mismatch in exchange for avoiding the increased experimental and computational burden of further model expansion.
\end{enumerate}
The choice between these strategies represents a critical trade-off between the accuracy of the final optimized solution and the practical costs of characterization.

\subsection{Block coordinate descent algorithm}

This section details the optimization framework for qubit frequency allocation, which we formulate as the following problem:
\begin{equation}
    \min_{f_i\in \mathcal{F}_i } G(f_1,f_2, \dots ,f_N)
    \label{eq:opt}
\end{equation}
Here, $\mathbf{f} = (f_1, f_2, \dots, f_N)$ is the vector of tunable device parameters, such as qubit frequencies, confined to a feasible set $\mathcal{F} = \prod_{i=1}^N \mathcal{F}_i$, where each $\mathcal{F}_i \subseteq \mathbb{R}$ is a closed, convex set. The objective function $G: \mathcal{F} \to \mathbb{R}$ is a continuous function that quantifies the performance of the quantum device as measured by the relevant calibration experiments. Crucially, $G(\mathbf{f})$ is not an analytical expression but a value estimated from the outcomes of physical calibration experiments, which are subject to inherent experimental noise.

The experimental nature of evaluating $G(\mathbf{f})$ makes gradient-based optimization methods impractical. Estimating the partial derivative $\frac{\partial G}{\partial f_i}$ using methods like finite differences would require numerous high-precision experiments for each parameter, rendering the process prohibitively slow and resource-intensive. This challenge motivates our use of a derivative-free optimization strategy. Specifically, we employ a Block Coordinate Descent (BCD) algorithm, which iteratively optimizes the system by solving smaller, more manageable subproblems.

To exploit the problem's structure, we partition the parameters into $M$ (possibly overlapping) blocks $\mathcal{B}=\{B_j\}_{j=1}^M$, where $\cup_j B_j=(f_1,f_2,...,f_N)$. In each step of our Block Coordinate Descent (BCD) algorithm, we solve a subproblem focused on a single block $B_j$. This subproblem is formulated as:
\begin{equation}
\min_{f_{j_i}\in \mathcal{F}_{j_i}} G_{B_j}(f_{j_1},f_{j_2}, \dots )
\end{equation}
where the minimization is performed over all parameters $\{f_{j_i}\}$ that constitute the block $B_j$. The cost function $G_{B_j}$ is defined based on a heuristic experimental model, $G'$, as:
\begin{equation}
G_{B_j}(u):=G'(u,\setminus B_j)
\end{equation}
In this definition, $u$ represents the set of parameters within block $B_j$, while $\setminus B_j$ represents the complementary set of parameters held constant during the subproblem optimization. The function $G':\Pi_{i=1}^N\mathcal{F}_i\rightarrow{}\mathbb{R}$ represents the value measured from a specific calibration experiment. The ideal choice, $G'=G$, would require a full-chip experiment for every subproblem, which is computationally prohibitive as the number of qubits increases. Therefore, by assuming that crosstalk effects are primarily local, we can adopt a more prudent approach: we define and select a tractable $G'$ that models a smaller, localized experiment sufficient for optimizing the parameters in block $B_j$.

A single iteration of the Block Coordinate Descent (BCD) algorithm involves selecting one block $B_j$ from the set of blocks $\mathcal{B}$ and solving the corresponding optimization subproblem. This subproblem aims to find the optimal parameters for the selected block, denoted $B_j^*$, by minimizing the local heuristic objective function $G_{B_j}$: 
\begin{equation}
    B_{j}^*\in \mathop{\arg\min}\limits_{B_{j}} G_{B_j}(B_{j})
    \label{eq:sub-opt}
\end{equation}
In this expression, the minimization $\arg\min_{B_j}$ is performed with respect to the set of parameters contained within block $B_j$, while all other parameters outside this block are held constant. After finding a solution vector $B_j^*=(f_{j_1}^*,f_{j_2}^*, \dots)$, the global parameter vector $f=(f_1,f_2, \dots ,f_N)$ is updated by replacing the parameters in $B_j$ with their new values from $B_j^*$. A complete pass through all blocks in $\mathcal{B}$ constitutes one epoch of the algorithm. The full pseudocode for this procedure is presented in Alg.~\ref{alg:bcd}.

\begin{algorithm}[H]
  \caption{Block Coordinate Descent}
  \label{alg:bcd}
  \begin{algorithmic}[1]
    \Require
      $f^{0,0}$ starting point;
      $\mathcal{F}$ parameter space;
      $\{B_1,B_2,\dots,B_M\}$ blocks of parameters;
      $G$ overall cost function;
      $\{G_1,G_2,\dots,G_M\}$ sub-cost functions
    \Ensure
      $G^*$, the convergence value of $G$
    \State set epoch $j \gets 0$
    \Repeat
      \State $j \gets j+1$
      \State set $f^{j,0} \gets f^{j-1,k}$
      \State set step $k \gets 0$
      \Repeat
        \State $k \gets k+1$
        \State choose $j_k \in \{1,2,\dots,M\}$ to determine block $B_{j_k}$
        \State choose $B_{j_k}^{j,k} \in 
          \operatorname*{arg\,min}\limits_{B_{j_k}}
          G_{B_{j_k}}\!\bigl(B_{j_k}\mid f^{j,k-1}\setminus B_{j_k}^{j,k-1}\bigr)$
        \State \label{line:update} $f^{j,k} \gets \bigl(B_{j_k}^{j,k},\, f^{j,k-1}\setminus B_{j_k}^{j,k-1}\bigr)$
      \Until{each parameter is optimized at least once}
    \Until{convergence}
  \end{algorithmic}
\end{algorithm}

There are three key components in implementing the BCD algorithm: the block strategy, how to select block each step, and how to optimize each subproblem.

For the block strategy, since we know the spatial distribution of qubits on superconducting quantum chips, 
we typically group spatially neighboring parameters together. 
Under the assumption of local crosstalk, the influence ranges of parameters within this block overlap. 
This implies that for a fixed number of parameters in a block, the influence range is minimized. 
A minimized influence range means we can more easily obtain the objective function value through calibration experiments. 
Therefore, this paper employs a non-overlapping block strategy centered around each qubit to narrow the search space for partitioning. 
Section \ref{sec:NNA} explores more suitable partitioning strategies within this space. Clearly, another parallel partitioning strategy space exists. 
When parameters within each grouping are spaced as far apart as possible in the parameter space, their influence ranges become mutually independent due to the assumption of local crosstalk. In this case, optimizing parameters independently within each region significantly reduces both the search space and the complexity of the objective function. However, this scenario is uncommon at current qubit counts and is therefore not discussed in detail in this paper.

Two related but distinct design choices underlie our framework. The \emph{optimization-block partition} determines how tunable parameters are grouped into blocks \(B_j\), thereby fixing the local search dimension \(|B_j|\). The \emph{reduced-objective footprint} determines how many error contributions must be retained when evaluating the block-local objective \(G_{B_j}\), and its size \(|C_j|\) controls the effective evaluation cost. Although the two are linked through the crosstalk graph, they are not identical.

Block size involves a direct trade-off. Larger blocks internalize more crosstalk within each subproblem, improving local faithfulness, but also inflate both the search cost (which grows as \(3^{|B_j|}\) or \(k^{|B_j|}\)) and the footprint needed for a faithful reduced objective. Conversely, very small blocks---in the extreme, single parameters---shrink each local search but weaken the sequence dependence that makes the SD-TSP ordering meaningful. The qubit-centered non-overlapping partition adopted here is therefore a practical compromise between optimization fidelity, reduced-objective locality, and tractable per-epoch cost, rather than a universally optimal rule.

For each block selection, the BCD algorithm can either adopt a fixed selection order and iterate, or choose based on certain rules each time. For instance, selecting the block direction that maximizes the reduction in the objective function is a common approach in the BCD algorithm. However, this can lead to issues in the present optimization problem, specifically the aforementioned difficulty in differentiation for this setting. Therefore, selecting a fixed order is a viable option for chip calibration problems. Furthermore, by fixing the optimization problem to be solved at each epoch, we can determine the computational complexity of the optimization algorithm at each step. This insight guides us toward developing more effective block strategies.

Finally, we address the optimization problem for the subproblems. Here, we have significantly reduced the dimensionality of the search space, making the choice of optimization algorithms relatively arbitrary. The only requirement is that each optimization algorithm provides its computational complexity so that the overall complexity of each epoch of the BCD algorithm can be calculated. Here we present a method below.

Solving the subproblem in Eq.~(\ref{eq:sub-opt}) is itself a multi-parameter optimization challenge of dimension $|B_j|$, which must be addressed without access to analytical gradients. To this end, we employ an iterative local search algorithm. Starting from the current parameters for the block, $B_j^0$, the method generates a sequence of updates $B_j^{t+1}=B_j^t+d_t r_n$. The step size $r_n$ is set according to the main BCD epoch number $n$; inspired by the stochastic approximation literature \cite{Lyu2020}, we use the diminishing step size $r_n=\frac{1}{n}$. The direction vector $d_t$ is determined at each sub-iteration by performing a search over a discrete set of local directions:
\begin{equation}
    d_t=\mathop{\arg\min}\limits_{d\in S} G_{B_j}(B_j+dr_n)
    \label{eq:sub-complexity}
\end{equation}
The search space $S$ is the set of all vectors whose components are restricted to $\{-1, 0, 1\}$:
\begin{equation}
    S=\{\mathbf{v}|v_i \in \{-1,0,1\},\forall i \in \{1,2,\dots,|B_j|\}\}.
\end{equation}
This procedure effectively evaluates the objective function at all $3^{|B_j|}$ points on a hypergrid centered at $B_j^t$. If a candidate point $B_j^t+dr_n$ lies outside the feasible domain where $G_{B_j}$ is defined, it is excluded from the minimization. Since the zero vector is included in $S$, this process guarantees that the objective function is non-increasing, as $G_{B_j}(B_j^{t+1}) \le G_{B_j}(B_j^t)$.


The primary cost of this procedure is the number of physical experiments required. We can quantify the experimental complexity for optimizing block $B_j$ as: 
\begin{equation}
    \mathcal{T}(B_j)=S_j\times T(G_{B_j})
\end{equation}
Here, $T(G_{B_j})$ represents the cost (e.g., time) of performing a single experiment to evaluate the local objective function $G_{B_j}$, and $S_j$ is the total number of such evaluations needed to solve the subproblem. This number of evaluations is determined by:
\begin{equation}
    S_j=S\times 3^{|B_j|}
\end{equation}
In this expression, $3^{|B_j|}$ is the size of the search-direction set $S$ defined previously. The variable $S$ in this specific context denotes the number of iterations performed by the iterative search in Eq.~(\ref{eq:sub-complexity}) to converge to a solution for the block.

\subsection{Convergence of BCD Algorithm}\label{sec:32}

This section analyzes the convergence properties of the BCD algorithm under conditions relevant to the qubit frequency allocation problem studied here. We begin by establishing convergence in an idealized theoretical setting before examining how practical factors like inexact optimization, measurement noise, and model mismatch affect performance.
To demonstrate convergence, we first make several standard assumptions about the objective function and the parameter space.
Under these conditions, we can guarantee that the value of the objective function converges.

\begin{assumption}
(Regularity)
\begin{itemize}
    \item The global objective function $G:\Pi_{i=1}^N\mathcal{F}_i\rightarrow{}\mathbb{R}$ is proper, lower bounded, and continuously differentiable on an open set containing $\Pi_{i=1}^N\mathcal{F}_i$.
    \item Each parameter space $F_i \subseteq \mathbb{R}$ is a closed, convex set.
\end{itemize}
\label{assu:reg}
\end{assumption}

\begin{assumption}
(Physical condition)

\begin{itemize}
    \item (Local crosstalk)For any block $B_j \in \mathcal{B}$, a local objective function $G_{B_i}$ can be defined such that its evaluation complexity $T(G_{B_i})$ is independent of the total number of qubits $N$.
    \item (Equivalence) The local objective function $G_{B_j}$ is a valid surrogate function for the global objective function $G$, meaning $G_{B_j}$ preserves the ordering induced by $G$ when all other parameters are held constant. This is the condition expressed in Eq.~(\ref{eq:optimal}).
\end{itemize}

\label{assu:phy}
\end{assumption}

\begin{theorem}
Assume that Assumptions 1 and 2 are satisfied. For the sequence $\{f^{t}\}$ generated by BCD, the sequence $\{G(f^t)\}$ converge to $G^*$ and there exists $f^*\in \Pi_{i=1}^N\mathcal{F}_i$ such that 
\begin{equation}
    G(f^*)=G^*
\end{equation}
\end{theorem}

Proof:
According to the definition of Eq.~(\ref{eq:sub-opt}), in this optimization process, we optimized from the initial parameter $B_{j_k}^{j,k-1}$ to the result $B_{j_k}^{j,k}$, thus establishing the order relation 
\[
    G_{B_{j_k}}(B_{j_k}^{j,k}|f^{j,k-1}\setminus B_{j_k}^{j,k-1})
    \leq 
    G_{B_{j_k}}(B_{j_k}^{j,k-1}|f^{j,k-1}\setminus B_{j_k}^{j,k-1})
\]

Then we known that function $h$ in Eq.~(\ref{eq:optimal}) is strictly increasing, it preserve the order relation, then we have 
\[
    G(B_{j_k}^{j,k}|f^{j,k-1}\setminus B_{j_k}^{j,k-1})
    \leq 
    G(B_{j_k}^{j,k-1}|f^{j,k-1}\setminus B_{j_k}^{j,k-1})
    \label{eq:improve-local}
\]
Due to the parameter update rules Line.~(\ref{line:update}) in Alg.~(\ref{alg:bcd}), we have the relationship
\[
    G(f^{j,k})
    =G(B_{j_k}^{j,k}|f^{j,k-1}\setminus B_{j_k}^{j,k-1})
    \leq 
    G(B_{j_k}^{j,k-1}|f^{j,k-1}\setminus B_{j_k}^{j,k-1})
    =G(f^{j,k-1})
    \label{eq:improve-global}
\]
i.e. each step of updating does not increase the objective value.

Because $G$ has lower bound $G\geq0$,
the non-increasing sequence $G(f^{j,k})$ converges to $G^*$.

For the sequence $\{f^{j,k}\}$ in closed set $\Pi_{i=1}^N\mathcal{F}_i$,
according to Bolzano–Weierstrass theorem, there exists subsequence $\{f^{t_m}\}$ converge to $f^*\in \Pi_{i=1}^N\mathcal{F}_i$ 
and the subsequence $G(f^{t_m})$ converge to $G^*$.
Then 
\[
    G(f^*)=G^*
\]

This completes the proof.\qedsymbol

Note that Theorem 1 merely demonstrates that the objective function value in the BCD algorithm can converge to a specific value where the objective function is defined. However, under the current assumptions, there are no constraints on the found parameter $f^*$, which is not necessarily a local optimum and may not even be a stationary point.\cite{Powell1973}. The result thus indicates that the BCD algorithm can only partially optimize the qubit frequency allocation problem. Of course, considering practical applications, BCD partially mitigates crosstalk phenomena in quantum chips through calibration experiments.\cite{Google2024}

If we wish to determine with certainty whether our optimization results for qubit frequency allocation represent at least one stationary point, we need to introduce additional strict assumptions about the objective function.

\begin{assumption}
(Stationary point convergence)
\begin{itemize}
    \item (KL property) The objective function $G$ satisfies the Kurdyka-Lojasiewicz (KL) property on its domain $\mathcal{F}$.
    \item (Unique solution) Each subproblem in Eq.~(\ref{eq:sub-opt}) admits a unique solution, and the objective function strictly decreases with each update.
\end{itemize}
\label{assu:sta}
\end{assumption}

When Assumption~\ref{assu:reg} and Assumption~\ref{assu:sta} are satisfied, the sequence generated by the block coordinate descent algorithm is guaranteed to converge to a stationary point of $G$ \cite{Tseng2001,Attouch2013}.

In practice, however, it is impossible to solve the subproblem in Eq.~(\ref{eq:sub-complexity}) with perfect precision. We terminate the local search after a finite number of iterations, $S$, which means the solution will inevitably contain some deviation from a true local minimum $B^*_j$. This inexactness can be bounded, for instance, by the following condition on the obtained solution $B_j^S$:
\begin{equation}
    \|B_j^* - B_j^S\|_{\infty} < \frac{1}{S}
\end{equation}
This condition characterizes the result of a single inexact block update. For the overall BCD algorithm, this implies that the sequence of generated parameters is produced by subproblem solutions that are consistently approximate. Therefore, the sequence of solutions $\{B_j^S\}$ obtained throughout the algorithm's execution must satisfy this bound for every block coordinate:
\begin{equation}
    \|B_j^* - B_j^S\|_{\infty} < \frac{1}{S} \quad \forall j=1,2,\dots
\end{equation}

A second, and more critical, challenge arises when the validity condition in Eq.~(\ref{eq:optimal}) is only approximately satisfied. When this occurs, the global cost function $G$ is no longer guaranteed to be non-increasing at each step, which can impede convergence. There are two primary reasons for this discrepancy. First, random measurement errors are introduced during the process of obtaining function values; the objective function used in optimization is not the ideal $G$ but rather a noisy version $\tilde{G} = G + \text{noise}$. Second, the effects of non-local crosstalk may be neglected if the heuristic model is incomplete, introducing a systematic error even with precise measurements. To investigate the impact of each of these factors separately, we conducted two sets of numerical experiments.

\subsection{Error simulator and simulation protocol} \label{sec:simulator}

\begin{figure}[t]
    \centering
    \includegraphics[width=1.0\linewidth]{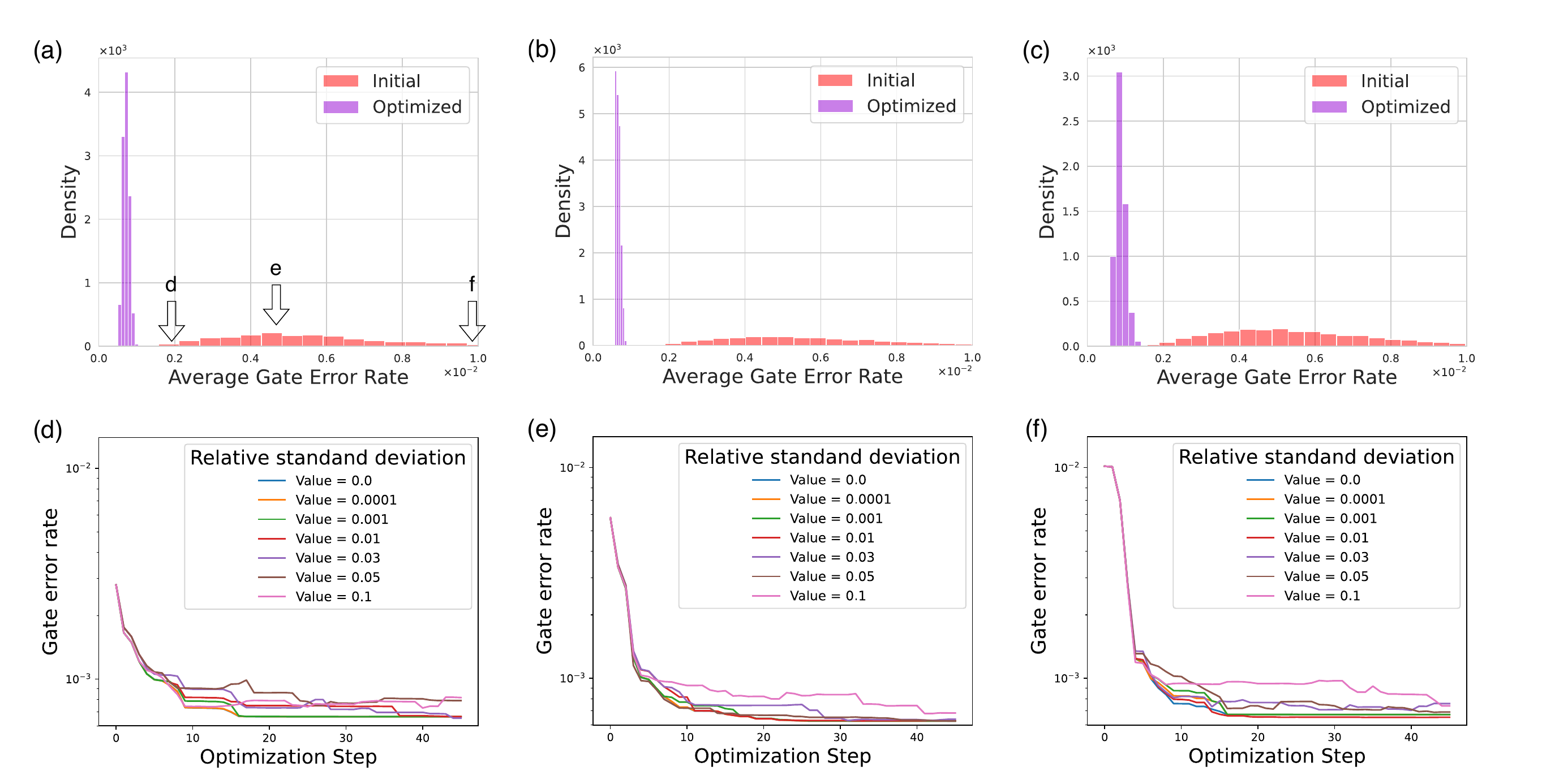}
    
    \caption{\textbf{Optimization performance.} 
    All panels correspond to the fixed-size \(3\times 3\) layout with 9 qubits and 21 tunable frequency-related parameters (9 single-qubit parameters and 12 two-qubit-gate parameters).
    (a)We employ probability density functions to describe the distribution of our optimization metric (the average gate error rate) before and after optimization with BCD-NNA, 
    starting from randomly chosen initial points under different simulator instances. 
    Here, each simulator instance is generated by randomly sampling the parameters of the physics-motivated error simulator within fixed ranges.
    (b,c) For a fixed simulator instance, distributions of the average gate error rate before and after optimization when starting from different randomly selected initial points. 
    During optimization in panel (c), the objective-function evaluations include a relative standard deviation of 0.2, whereas no such variability is present in panel (b). 
    (d,e,f) Demonstration of the optimization process showing the average gate error rate evaluated from the underlying noiseless simulator varies with the number of optimization steps starting from different initial points in panel (a). 
    The curves show the noiseless average gate error rate, while the optimization itself uses evaluations with different relative standard deviations.
    }
    \label{fig:optimization-performance}
\end{figure}

To make the numerical experiments more explicit, we now summarize the error simulator and the simulation protocol used in Fig.~\ref{fig:optimization-performance}, Fig.~\ref{fig:nonlocal-crosstalk}, and Fig.~\ref{fig:nna-bcd}. The simulator is designed as a physics-motivated surrogate objective for qubit frequency allocation on superconducting quantum processors. Rather than presenting implementation details, we formulate it at the level of physically interpretable contributions. For the fixed-qubit-number experiments reported in these figures, we consider a \(3\times 3\) qubit layout containing 9 qubits. In this setting, the optimization variable contains 21 frequency-related parameters in total: 9 single-qubit frequency parameters and 12 two-qubit-gate frequency parameters associated with the nearest-neighbor couplings of the \(3\times 3\) lattice.

For a frequency configuration \(f=(f_1,\dots,f_N,\dots)\), where the relevant variables include qubit idle frequencies and two-qubit interaction frequencies, we define a circuit-level error estimator
\begin{equation}
    \mathcal{E}_{\mathrm{circ}}(f)=\mathcal{E}_{\mathrm{1b}}(f)+\mathcal{E}_{\mathrm{NN}}(f)+\mathcal{E}_{\mathrm{NL}}(f),
    \label{eq:error-decomp-main}
\end{equation}
where \(\mathcal{E}_{\mathrm{1b}}\) collects single-body contributions, \(\mathcal{E}_{\mathrm{NN}}\) collects topology-local multi-qubit contributions, and \(\mathcal{E}_{\mathrm{NL}}\) represents nonlocal multi-qubit corrections used to test model mismatch. More explicitly,
\begin{equation}
    \mathcal{E}_{\mathrm{1b}}(f)=
    \sum_{i\in \mathcal{Q}} \mathcal{E}_i^{(\phi)}(f_i)
    +\sum_{i\in \mathcal{Q}} \mathcal{E}_i^{(T_1)}(f_i),
\end{equation}
\begin{equation}
    \mathcal{E}_{\mathrm{NN}}(f)=
    \sum_{\langle i,j\rangle \in \mathcal{E}} \mathcal{E}_{ij}^{(\mathrm{coll})}(f_i,f_j)
    +\sum_{\langle i,j\rangle \in \mathcal{E}} \mathcal{E}_{ij}^{(\mathrm{2q},\phi)}(f)
    +\sum_{\langle i,j\rangle \in \mathcal{E}} \mathcal{E}_{ij}^{(\mathrm{2q},T_1)}(f)
    +\sum_{\langle i,j\rangle \in \mathcal{E}} \mathcal{E}_{ij}^{(\mathrm{fluxXT})}(f),
\end{equation}
\begin{equation}
    \mathcal{E}_{\mathrm{NL}}(f)=
    \sum_{(i,j)\in \mathcal{P}_{\mathrm{far}}} \mathcal{E}_{ij}^{(\mathrm{far})}(f),
\end{equation}
with \(\mathcal{Q}\) the qubit set, \(\mathcal{E}\) the set of nearest-neighbor couplings on the chip graph, and \(\mathcal{P}_{\mathrm{far}}\) a set of non-neighbor qubit or gate pairs used only when testing the effect of nonlocal crosstalk. This decomposition highlights the distinction between on-site terms, nearest-neighbor many-body terms, and non-nearest-neighbor corrections.

The single-body terms model the dominant one-qubit frequency dependence: \(\mathcal{E}_i^{(\phi)}\) describes dephasing-related error and \(\mathcal{E}_i^{(T_1)}\) describes relaxation-related error. The nearest-neighbor multi-body terms capture the fact that gate performance depends not only on each qubit individually but also on nearby simultaneously operated elements. In particular, \(\mathcal{E}_{ij}^{(\mathrm{coll})}\) penalizes frequency collisions or near-collisions between coupled elements, \(\mathcal{E}_{ij}^{(\mathrm{2q},\phi)}\) and \(\mathcal{E}_{ij}^{(\mathrm{2q},T_1)}\) describe dephasing and relaxation accumulated along two-qubit gate trajectories, and \(\mathcal{E}_{ij}^{(\mathrm{fluxXT})}\) models flux-crosstalk-induced shifts of the effective qubit frequencies. The optional term \(\mathcal{E}_{\mathrm{NL}}\) is introduced only in the robustness tests of Fig.~\ref{fig:nonlocal-crosstalk}, where we intentionally compare optimization with and without accounting for nonlocal crosstalk.

In the simulations, ``random error model'' does not mean that noise is added arbitrarily to the final result. Instead, for each independent trial we first sample the coefficients and characteristic scales appearing in the above terms from fixed physically motivated intervals, thereby generating one static device instance. The optimizer is then run on that instance. By contrast, ``stochastic noise'' refers to measurement uncertainty in the objective-function evaluation during optimization: the optimizer does not access \(\mathcal{E}_{\mathrm{circ}}(f)\) directly, but rather a noisy estimate of it.

More specifically, if \(\mathcal{E}_{\mathrm{circ}}(f)\) denotes the underlying noiseless circuit-level error estimated by the simulator, then the value provided to the optimizer is
\begin{equation}
    \tilde{\mathcal{E}}_{\mathrm{circ}}(f)=\mathcal{E}_{\mathrm{circ}}(f)\bigl(1+\xi\bigr)+\delta_{\mathrm{miss}}(f),
    \label{eq:noisy-eval-main}
\end{equation}
where \(\xi\) is a zero-mean random variable describing finite-shot or measurement-like uncertainty, and \(\delta_{\mathrm{miss}}(f)\) is a systematic model-mismatch term that appears when the optimizer ignores nonlocal crosstalk present in the underlying simulator. In our numerical implementation, the random factor is sampled independently at each function evaluation, so it perturbs the optimization trajectory rather than merely the final reported point.

The quantity directly optimized in the numerical study is therefore the noisy circuit-level objective \(\tilde{\mathcal{E}}_{\mathrm{circ}}(f)\). However, the quantity plotted in Fig.~\ref{fig:optimization-performance}, Fig.~\ref{fig:nonlocal-crosstalk}, and Fig.~\ref{fig:nna-bcd} is the corresponding \emph{average gate error rate}, obtained from the underlying noiseless circuit-level error \(\mathcal{E}_{\mathrm{circ}}(f)\). If the circuit template used for evaluation contains \(G\) effective gate operations and \(P_{\mathrm{circ}}(f)=1-\mathcal{E}_{\mathrm{circ}}(f)\) denotes the circuit success probability, then we define the average gate success probability \(\bar{p}_g(f)\) by
\begin{equation}
    P_{\mathrm{circ}}(f)=\bar{p}_g(f)^G,
    \qquad
    \bar{p}_g(f)=\bigl(1-\mathcal{E}_{\mathrm{circ}}(f)\bigr)^{1/G},
\end{equation}
and hence the reported average gate error rate is
\begin{equation}
    \bar{\varepsilon}_g(f)=1-\bigl(1-\mathcal{E}_{\mathrm{circ}}(f)\bigr)^{1/G}.
    \label{eq:avg-gate-error-main}
\end{equation}
Thus, although the simulator is constructed and optimized in terms of a summed circuit-level error estimator, the final metric reported in the figures is the effective average gate error rate inferred from the corresponding circuit success probability through a product formula for gate success probabilities.

Finally, the distributions shown in Fig.~\ref{fig:optimization-performance}a should not be interpreted as distributions generated solely by additive measurement noise around one fixed optimum, so a Gaussian form is not expected in general. Each histogram or density plot mixes contributions from different randomly sampled device instances, different random initial points, and the nonlinear action of the optimizer on a nonconvex landscape. Even when the evaluation noise itself is sampled from a simple bounded distribution, the resulting distribution of optimized average gate error rates is shaped primarily by the geometry of the optimization landscape and basin structure, rather than by a direct central-limit mechanism.
We also note that the efficiency comparisons in later figures are reported in terms of the algorithmic cost proxies defined in Sec.~\ref{sec:complexity}, not wall-clock runtime; this keeps the ordering comparison independent of a particular software or hardware environment.

\subsection{Performance of BCD Algorithm}

\begin{figure}[t]
    \centering
    \includegraphics[width=1.0\linewidth]{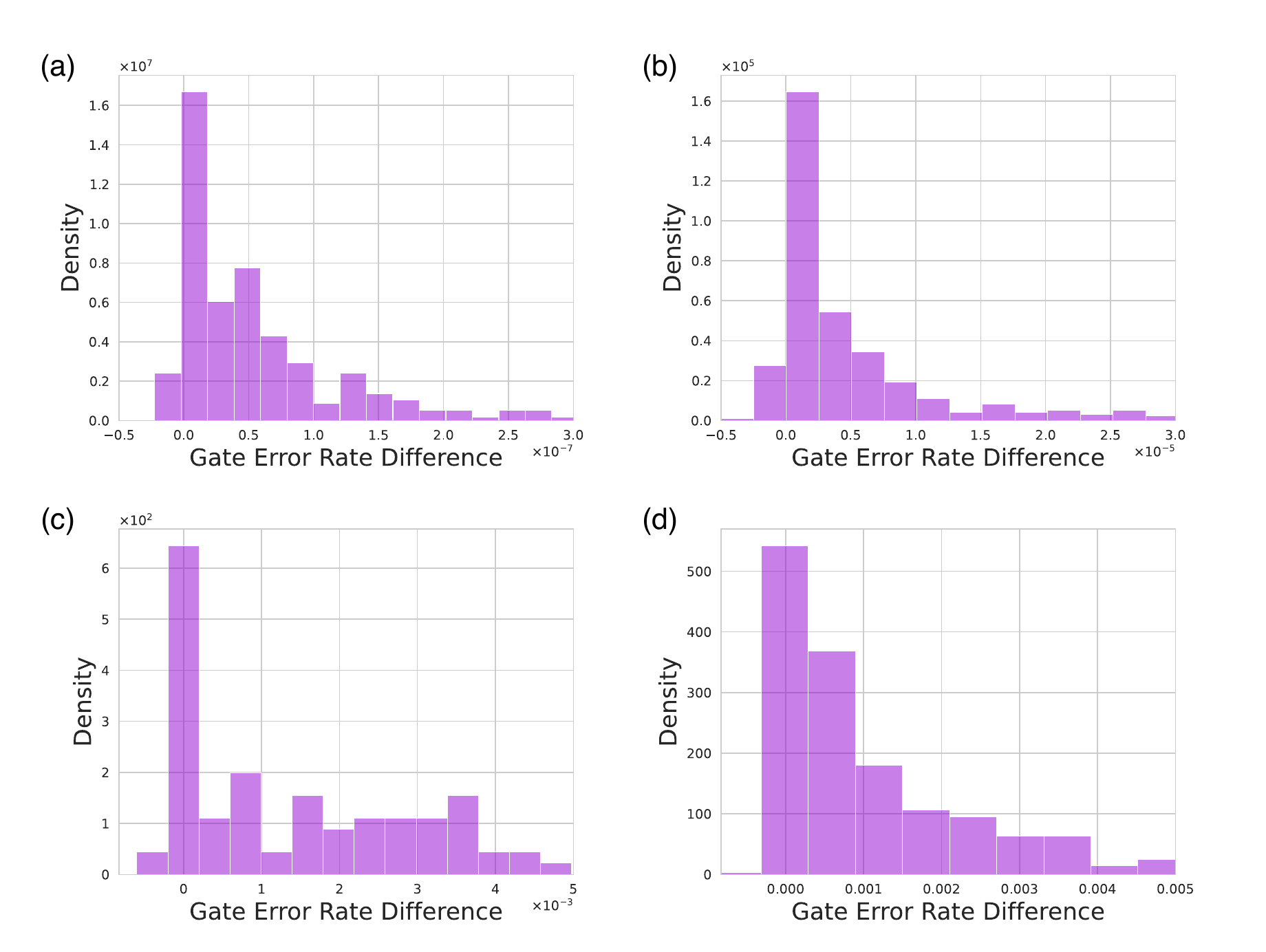}
    \caption{\textbf{Impact of nonlocal crosstalk errors on the optimization process.}
    The simulations in all panels use the same fixed \(3\times 3\) layout with 9 qubits and 21 tunable frequency-related parameters.
    In the numerical simulation, we altered the maximum range of random selection for nonlocal crosstalk parameters in the error simulator, incrementally increasing it from (a) to (d), while maintaining the random selection ranges for other local crosstalk parameters unchanged.
    For each randomly generated simulator instance, we then compare optimization using a local crosstalk model with optimization using a model that explicitly includes the nonlocal crosstalk contribution.
    The probability density plots show the distribution of differences in average gate error rates, calculated as the results from crosstalk-unaware optimization minus those from crosstalk-aware optimization for the final noiseless reported metric.}
    \label{fig:nonlocal-crosstalk}
\end{figure}

To evaluate the performance and robustness of the BCD algorithm under realistic conditions, we conduct a series of numerical simulations based on the error simulator introduced above. For the fixed-qubit-number experiments shown in Fig.~\ref{fig:optimization-performance}, Fig.~\ref{fig:nonlocal-crosstalk}, and Fig.~\ref{fig:nna-bcd}, we use a \(3\times 3\) processor topology with 9 qubits. The corresponding optimization problem contains 21 tunable frequency-related parameters, including 9 single-qubit frequency parameters and 12 two-qubit-gate frequency parameters. By contrast, the scaling study in Fig.~\ref{fig:scaling} varies the number of qubits and is used specifically to examine how optimization quality and complexity change with system size. These simulations probe three distinct sources of variation: differences between device instances, sensitivity to initialization in a nonconvex landscape, and stochastic uncertainty in objective-function evaluations.

First, different simulated devices are generated by randomly sampling the coefficients entering the single-body, nearest-neighbor, and, when applicable, nonlocal terms in Eq.~(\ref{eq:error-decomp-main}) within fixed physically motivated ranges. Thus, ``random error model'' in Fig.~\ref{fig:optimization-performance}a means a random draw of one full simulator instance, not a random perturbation of an already optimized result. This allows us to assess how broadly the optimization strategy performs across varying hardware conditions.

Second, even for a fixed simulator instance, the optimization problem remains nonconvex. We therefore test sensitivity to initialization by repeating the optimization from multiple random initial frequency configurations. In the figures, ``random initial points'' refer to these independently sampled starting frequency allocations within the feasible frequency window.

Third, to model the fact that an experimental objective function is inferred from measurements rather than evaluated exactly, the optimizer is supplied with the noisy circuit-level estimate \(\tilde{\mathcal{E}}_{\mathrm{circ}}(f)\) in Eq.~(\ref{eq:noisy-eval-main}) instead of the exact simulator value \(\mathcal{E}_{\mathrm{circ}}(f)\). This noise is introduced at the stage of objective-function query during the optimization loop, so it affects the trajectory of the optimizer itself. By contrast, the final reported performance in the plots is always the noiseless average gate error rate \(\bar{\varepsilon}_g(f)\) defined by Eq.~(\ref{eq:avg-gate-error-main}), which is obtained from the underlying noiseless circuit-level error estimator.
The results of these investigations are summarized in Fig.~\ref{fig:optimization-performance}. Under the local-crosstalk assumption, the BCD algorithm consistently improves system performance across all tested error models and from all random initializations. The non-convex nature of the problem is apparent, as different starting points lead to final values of varying quality. Nonetheless, even the worst-performing optimization run achieves a significant improvement over its random initial state. When stochastic measurement noise is introduced, the final performance generally degrades. Interestingly, the noise can occasionally have a beneficial annealing-like effect, enabling the optimizer to escape a shallow local minimum and converge to a superior solution than what might be achieved in a noiseless scenario.

The preceding simulations operated under the assumption that the error model perfectly matched the simulated physical reality (i.e., both contained only local crosstalk). We now consider the more complex case of model mismatch, where the physical system exhibits non-local crosstalk, but the optimization algorithm proceeds with a model that only accounts for local effects. This mismatch introduces a systematic error into the optimization process. As illustrated in Fig.~\ref{fig:nonlocal-crosstalk}, we compare two strategies in this scenario: 1) using a more complex algorithm that correctly models the non-local effects, and 2) using the simpler, misspecified algorithm based on the local crosstalk assumption.

The results show that employing the accurate, more complex model yields significantly better optimization outcomes. While the simpler, misspecified model still provides some improvement, its effectiveness is fundamentally limited by the systematic error. This highlights a critical trade-off: the decision to incorporate non-local effects into the optimization model depends on the magnitude of these effects versus the increased complexity and computational cost of the corresponding algorithm.

In summary, the practical application of this optimization framework involves navigating a trade-off between algorithmic stability, computational cost, and ultimate performance, which is dictated by how closely the idealized assumptions are met.

\begin{itemize}
    \item \textbf{Strict Validity and Model Accuracy:} When the crosstalk model is accurate and measurements are precise (approximating the strict satisfaction of the validity condition in Eq.~(\ref{eq:optimal})), the algorithm's stability is assured, and it can reliably converge, as demonstrated in the noiseless cases in Fig.~\ref{fig:optimization-performance}. Achieving this in practice requires either a system with negligible unmodeled effects or a significant investment in characterization and computation to build a highly accurate model.

    \item \textbf{Approximate Validity and Practical Constraints:} In practice, the validity condition is often only approximately satisfied due to factors like measurement noise or model mismatch. \textbf{Measurement noise}, as shown in Fig.~\ref{fig:optimization-performance}, introduces variance and can degrade final performance, although it may occasionally help escape local minima. \textbf{Model mismatch}, as shown in Fig.~\ref{fig:nonlocal-crosstalk}, imposes a more fundamental limitation on performance. Tolerating this systematic error leads to a simpler, more efficient algorithm but results in a suboptimal solution. Correcting for it improves the outcome at the cost of higher algorithmic complexity.
\end{itemize}

Ultimately, the choice of strategy depends on the specific requirements of the quantum system, balancing the need for performance against the available resources for characterization and computation.
Taken together, the numerical results in Fig.~\ref{fig:optimization-performance} and Fig.~\ref{fig:nonlocal-crosstalk} lead to three observations. Under the local-crosstalk assumption, the BCD-based framework consistently lowers the average gate error rate across both randomly sampled simulator instances and random initial points. Noisy objective-function evaluations degrade optimization quality only gradually, so the method remains useful under realistic measurement uncertainty. When the assumed local model omits nonlocal crosstalk that is actually present, the resulting performance loss is systematic but moderate for the mismatch levels tested here, which helps delineate the practical regime of the locality-based approximation. These simulations are not intended as claims of device-specific experimental performance; they provide a controlled test showing that the proposed optimization architecture offers a favorable trade-off among optimization gain, robustness, and algorithmic cost within a clearly specified surrogate model.

\subsection{Complexity of BCD}\label{sec:complexity}
Throughout this work ``complexity'' refers to an algorithmic cost proxy for one optimization epoch, not to hardware-dependent wall-clock runtime. The distinction matters because the central claim of the proposed ordering strategy is not that it improves every metric simultaneously, but that it achieves a more favorable cost--quality trade-off within the same BCD framework. Accordingly, the numerical comparisons below report both optimization quality and per-epoch algorithmic cost, so that the benefit of topology-aware ordering can be assessed in a balanced way.

The overall computational complexity of the Block Coordinate Descent (BCD) algorithm can be decomposed into three fundamental components:
(1) the number of epochs, in which all parameters are updated once;
(2) the number of cost function evaluations, denoted by $S_i$, that occur within each block during parameter optimization; and 
(3) the intrinsic computational cost of evaluating each cost function, represented by $T_{G_i}$. 
Nevertheless, under identical computational resources allocated per epoch, the performance improvement obtained per epoch typically exhibits a gradually diminishing trend. 
Hence, from a practical perspective, the total number of epochs is usually fixed to a specific value. 
Subsequently, the analysis of algorithmic complexity primarily focuses on the variation of the latter two components-namely $S_i$ and $T_{G_i}$.

In the BCD framework, the set of parameters is partitioned into multiple disjoint blocks, and the parameters within each block are optimized sequentially. 
The manner in which the parameters are grouped significantly influences the effective search space size, particularly in the presence of crosstalk structures among parameters. 
For instance, consider a system with $3N$ parameters that are divided into $M=N$ blocks. 
In this scenario, the average number of parameters contained in each block can be approximated as $| B_i |= 3$, implying that each block spans a three-dimensional search subspace. 
Assuming that each parameter can be selected from $k$ discrete values, the corresponding search space size for each block is given by:
\[
S_i = k^{| B_i |}
\]

In contrast, if all parameters are divided into a single block (i.e., $M=1$), then the BCD algorithm essentially loses its decompositional advantage, and the optimization becomes equivalent to a global search over the entire parameter space. 
Under such conditions, the dimensionality of the search space is $3N$, and the corresponding search size increases exponentially as $k^{3N}$.

\begin{figure}
    \centering
    \includegraphics[width=1.0\linewidth]{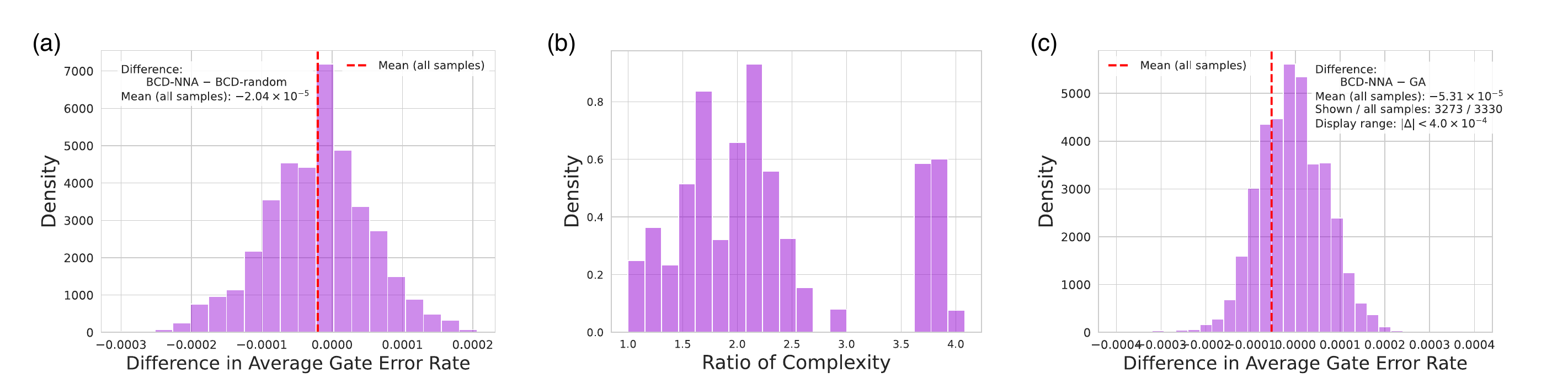}
    
    \caption{\textbf{NNA for BCD and comparison with a genetic-algorithm baseline.}
    All panels are obtained for the fixed \(3\times 3\) layout with 9 qubits and 21 tunable frequency-related parameters.
    (a) Probability density plot of the difference in average gate error rate, calculated as BCD-NNA minus BCD-random. Negative values indicate better final optimization quality for BCD-NNA. The distribution is centered close to zero, with a slightly negative sample mean, showing that the NNA ordering retains calibration quality while slightly outperforming random block orders on average.
    (b) Probability density plot of the empirical algorithmic cost ratio, calculated as BCD-random divided by BCD-NNA, demonstrating markedly lower complexity for BCD-NNA.
    (c) Probability density plot of the difference in average gate error rate, calculated as BCD-NNA minus GA, where GA denotes a genetic-algorithm baseline. Negative values indicate better final optimization quality for BCD-NNA. The distribution is again centered close to zero, with a slightly negative sample mean, indicating that BCD-NNA achieves optimization quality comparable to the GA baseline and is slightly better on average for the tested simulator instances.
    Here the cost is not wall-clock runtime; rather, it is the per-epoch algorithmic cost proxy defined by Eq.~(\ref{eq:complexity-empirical})--Eq.~(\ref{eq:complexity-empirical-total}), namely the sum over blocks of the local search cost multiplied by the effective reduced-objective evaluation cost.
    In panels (a) and (b), the difference between the compared methods arises solely from the block-update ordering strategy inside the same BCD framework. In panel (c), by contrast, BCD-NNA is compared with a population-based heuristic optimizer that does not explicitly use the topology-aware SD-TSP structure. Together, these panels show that the advantage of the proposed method comes from combining block-coordinate decomposition with a low-cost topology-aware ordering rule, rather than from heuristic search alone.
    }
    \label{fig:nna-bcd}
\end{figure}

It is important to note that the BCD algorithm operates in a stepwise optimization manner after the parameters are partitioned into blocks. 
Each resulting subproblem may still admit simplification by exploiting additional structural properties inherent to the specific problem. 
To characterize the upper-bound behavior of the algorithm, the complexity associated with a worst-case exhaustive verification is typically employed to estimate the maximum complexity of BCD. 
However, in practical implementations, strategies such as the \emph{diminishing radius} technique \cite{Lyu2020} are often introduced to substantially lower the effective search complexity from the naive exponential order of $k^n$ to a reduced order of $\mathcal{O}(3^n)$. 
Furthermore, various optimization algorithms may be adopted to further accelerate the solution process for each subproblem. 
It should be emphasized that the choice of specific optimization techniques inevitably leads to variations in the computational complexity. 
To characterize the sub-algorithm complexity more precisely in the present manuscript, we distinguish two complementary notions of algorithmic cost:

\begin{enumerate}
    \item \textbf{Empirical algorithmic cost proxy:} 
    For the implemented numerical procedure, the per-block cost is modeled as the product of the local search cost and the effective cost of one reduced-objective evaluation. Specifically, the inner derivative-free local search explores a discrete neighborhood of size \(3^{|B_i|}\) at each local step, as defined by Eq.~(\ref{eq:sub-complexity}), and the cost of one reduced-objective evaluation is represented by the number of retained simulator terms in that reduced local model, denoted \(N_{\mathrm{term}}(B_i)\). Therefore, for a fixed local iteration budget \(S\), we write
    \[
    \mathcal{C}^{\mathrm{emp}}_i
    =
    S\,3^{|B_i|}N_{\mathrm{term}}(B_i).
    \]
    Since \(S\) is fixed across the compared strategies, it acts as a common multiplicative factor, and the relative ordering cost is governed by \(3^{|B_i|}N_{\mathrm{term}}(B_i)\). This is the complexity notion used for the fixed-size numerical comparison of different traversal orders.
    
    \item \textbf{Search-space complexity model:} $S_i = k^{|B_i|}$ where $k \approx 100$, following the theoretical framework established in Ref.~\cite{Klimov2020}. In that work, each subproblem involves optimization over all possible frequency configurations for the gates in the block, yielding a search space of size $k^{|B_i|}$ where $k$ is the number of discrete frequency options per gate (typically $k \sim 10^2$). This model represents the worst-case complexity when exhaustive or grid search is required.
\end{enumerate}

The quantity \(N_{\mathrm{term}}(B_i)\) should be understood as a structured proxy for the cost of evaluating the reduced local objective, rather than as a literal count of hardware shots or processor runtime. In the physics-motivated simulator used here, the reduced objective for a block contains only those single-body, nearest-neighbor, and, where applicable, crosstalk-related contributions that fall within the reduced footprint associated with that block. Hence \(N_{\mathrm{term}}(B_i)\) increases when the chosen block ordering induces a larger effective local footprint, and decreases when the ordering keeps successive block objectives spatially compact. This is precisely why the block-ordering problem can be meaningfully optimized through the SD-TSP formulation.

The distinction between these two models has significant implications for evaluating the benefits of ordering optimization. Under the empirical model, which reflects our efficient implementation, the improvement from optimized block ordering is modest (Fig.~\ref{fig:nna-bcd}a). However, under the search-space model, which represents the theoretical upper bound applicable to less efficient optimization procedures or exhaustive verification, the improvement becomes substantially larger (Fig.~\ref{fig:nna-bcd}b). This observation highlights that the benefits of the proposed SD-TSP formulation scale with the complexity of the underlying sub-algorithm, making it particularly valuable for scenarios where exhaustive or near-exhaustive search is required.

For the theoretical analysis of scaling behavior, we adopt the search-space complexity model $S_i = k^{|B_i|}$ as the representative expression, consistent with the framework of Ref.~\cite{Klimov2020}.

After dividing the parameters in the BCD algorithm, 
it is necessary to consider not only the change of the search space, 
but also the complexity of obtaining the reduced cost function.
It is also necessary to consider the complexity of obtaining the cost objective function.
For example, the value of XEB experiment containing $\vert C_i \vert$ qubits is used as the reduced cost function.
To satisfy Eq.~(\ref{eq:optimal}), we can find that $\vert C_i \vert$ depends on $B_i$
This process requires multiple calculations of Eq.~(\ref{eq:xeb_complexity})
\begin{equation}
    p_U(x_i)=\vert <x_i \vert U \vert 0^{\vert C_i \vert} > \vert ^2
    \label{eq:xeb_complexity}
\end{equation}
If the Feynman path integral method is used,
the complexity of the computation increases exponentially with the number of contained qubits $\vert C_i \vert$.
Also the complexity of the computation increases exponentially with the depth of the circuit, 
though the depth of the circuit is usually fixed in this problem.
Considering also that the total number of measurement samples is $S$ times.
The computational complexity is denoted as $S\times t^{\vert C_i \vert}$.
Since in the process of calculating the XEB of the circuit, 
the time of classical computation will often be larger than the time of quantum circuit sampling.
The complexity of the classical computation is used as the complexity of the reduced objective function.
Then the complexity of getting cost function is $T_{G_i}=S\times t^{\vert C_i \vert}$

Of course, it is possible to use a more efficient way of calculating XEB or to use other output of the circuit as the reduced cost function, 
this complexity will be discussed as an example in this paper.
If XEB continues to be chosen as the reduced cost function,
it is necessary to reduce the number of qubits contained in the circuit as much as possible based on known crosstalk information.

Accordingly, the per-epoch algorithmic cost can be written in the generic form
\begin{equation}
    \mathcal{T}
    =
    \sum_{i=1}^{M}\mathcal{T}_i(B_i)
    =
    \sum_{i=1}^{M} S_i \times T_{G_i},
    \label{eq:complexity-generic}
\end{equation}

Under the empirical algorithmic cost proxy used in the implemented numerical simulations, we take
\begin{equation}
    S_i = S\,3^{|B_i|},
    \qquad
    T_{G_i}\propto N_{\mathrm{term}}(B_i),
    \label{eq:complexity-empirical}
\end{equation}
so that
\begin{equation}
    \mathcal{T}_{\mathrm{emp}}
    \propto
    \sum_{i=1}^{M} S\,3^{|B_i|}N_{\mathrm{term}}(B_i).
    \label{eq:complexity-empirical-total}
\end{equation}

By contrast, under the search-space model adopted for asymptotic scaling discussion, one may write
\begin{equation}
    \mathcal{T}_{\mathrm{search}}
    =
    \sum_{i=1}^{M} S\times k^{|B_i|}t^{|C_i|},
    \label{eq:complexity-search}
\end{equation}
where \(k^{|B_i|}\) represents the effective size of the block search space and \(t^{|C_i|}\) represents the growth of the reduced-objective evaluation cost with the reduced circuit footprint \(|C_i|\), as in the XEB-style example discussed below.

\begin{figure}
    \centering
    \includegraphics[width=1.0\linewidth]{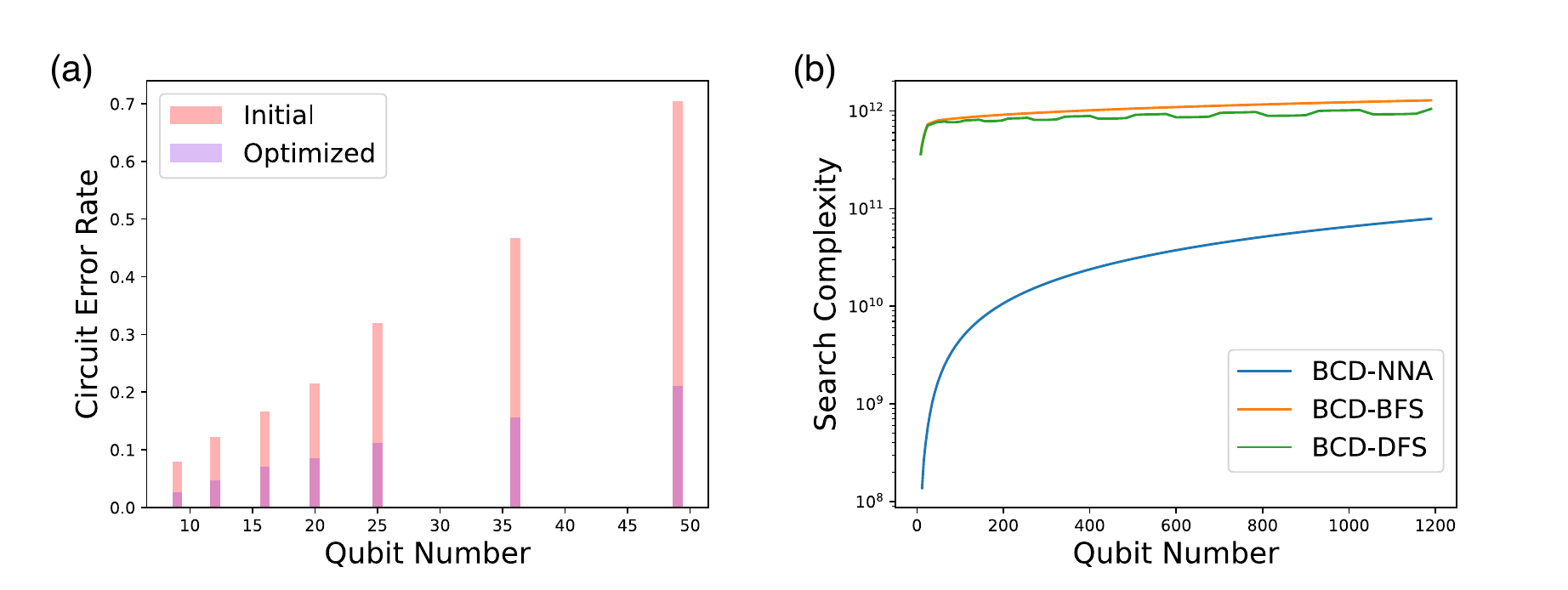}
    \caption{\textbf{Algorithm scaling with increasing number of qubits.}
    (a) Optimization of quantum chips with varying numbers of qubits using the BCD-NNA algorithm: comparison of circuit error rates before and after optimization. 
    (b) Algorithmic complexity comparison under the search-space complexity model ($S_i = k^{|B_i|}$ with $k=100$), following the theoretical framework of Ref.~\cite{Klimov2020}. This model represents the worst-case complexity for derivative-free optimization over all frequency configurations. It is used here to illustrate asymptotic scaling trends rather than implementation-specific runtime. Under this model, BCD-NNA achieves substantial complexity reduction compared to graph-based heuristics (BFS, DFS) employed in the original Snake implementation \cite{Google2024}, demonstrating the significant benefit of the SD-TSP formulation for optimizing block order.
    }
    \label{fig:scaling}
\end{figure}

These two complexity notions are used for different numerical purposes in the manuscript. For the fixed-size comparison of traversal strategies in Fig.~\ref{fig:nna-bcd}b, we report the ratio of empirical algorithmic cost proxies, because this quantity most directly reflects the relative cost of the implemented block-wise optimization procedure. For the system-size scaling comparison in Fig.~\ref{fig:scaling}b, we report the search-space complexity model, because it more transparently illustrates the asymptotic advantage of improved ordering when the underlying derivative-free subproblem becomes increasingly expensive.
Of course, the definition of algorithmic complexity varies depending on the specific optimization algorithms and the choice of different reduced objective function $G_i$. However, our primary concern lies in how the overall complexity evolves with the increasing number of qubits in the system under the assumption of local crosstalk. This evolution determines whether we can efficiently solve the qubit frequency allocation problem for larger-scale quantum chips.
When only local crosstalk exists in the system, 
with a given upper bound on the block size, 
the upper bound of function complexity ($T_{max}$) for each block does not change with the system size, and at the same time, the upper bound of search space ($S_{max}$) is also independent of the system size, 
which means that 
\begin{equation}
    \mathcal{T}
    =\sum_{i=1}^{M}\mathcal{T}_i
    =\sum_{i=1}^{M}S_i\times T_{G_{i}}
    \leq\sum_{i=1}^{M}S_{max}\times T_{max} 
    =M\times S_{max}\times T_{max} 
    \label{eq:complexity-upperbound}
\end{equation}

Here M is related to the total number of block in the system. 
If parameters are grouped around each qubit, this implies that $M=N$.
Therefore, when applying the BCD algorithm for one epoch of parameter optimization, the algorithm's complexity scales with the system size is $\mathcal{O}(N)$.

\subsection{Nearest-neighbor algorithm}\label{sec:NNA}

In selecting the parameter block strategy for the BCD algorithm, 
the exponential differences may exist between the complexity of different simplified calibration circuits, 
so in the process of optimizing different classes of parameters,
the complexity of the simplified calibration circuit is an important factor affecting the choice of the BCD algorithm.

When the complexity of the cost function is constant or polynomial with increasing parameters, 
it has been discussed\cite{Nutini2017}.
Here we focus on when the cost function increases polynomially or exponentially with the increase of the parameters,
how should the parameter block strategy be chosen for the BCD algorithm.
For superconducting quantum systems, it is clear that when the parameters divided into one block of the qubits are more dispersed, 
the complexity of obtaining the fidelity of the quantum circuit increases rapidly.
Even though the optimization task can be completed in one epoch by such a block strategy, 
the complexity of this strategy is most likely larger than other strategies.
Therefore, the main objective here is to find the BCD algorithm in one epoch with less complexity.

Let $G = (V, E)$ be a complete directed graph, where
\[
V = \{1, 2, \ldots, n\}
\]
is the set of cities, and $E = \{(i, j) \mid i, j \in V, i \neq j\}$ is the set of directed edges representing possible travel between cities.

A path (or Hamiltonian path) in $G$ is an ordering (permutation)
\[
\pi = (\pi_1, \pi_2, \ldots, \pi_n)
\]
of all vertices $V$, which represents the visiting order of the cities by a traveling agent.

Unlike the classical Traveling Salesman Problem (TSP\cite{Held1961,Lin1965,Dorigo1997,Helsgaun2000,Rego2011}), where the cost of traveling from city $i$ to city $j$ is fixed and denoted by $c(i,j)$, in the \emph{Traveling Salesman Problem with Path-Dependent Costs (TSP-PDC)}, the cost depends on both the current edge $(i,j)$ and the preceding history of the tour.

\noindent Formally, we define the path-dependent cost function as:
\[
C : V \times V \times \mathcal{H} \to \mathbb{R}_{\ge 0}
\]
where $\mathcal{H}$ denotes the set of all finite sequences of visited cities (i.e., partial tours), and
\[
C(i,j, H)
\]
represents the cost of traveling from city $i$ to city $j$ given that the sequence of previously visited cities (the current history) is $H = (v_1, v_2, \ldots, v_k)$.

\noindent The total cost of a complete tour $\pi = (\pi_1, \pi_2, \ldots, \pi_n)$ under a cost function $C$ is then defined recursively as:
\[
Cost(\pi; C) = \sum_{t=1}^{n-1} C\big(\pi_t, \pi_{t+1}, (\pi_1, \pi_2, \ldots, \pi_t)\big)
\]

\begin{figure*}
    \centering
    \includegraphics[width=1\linewidth]{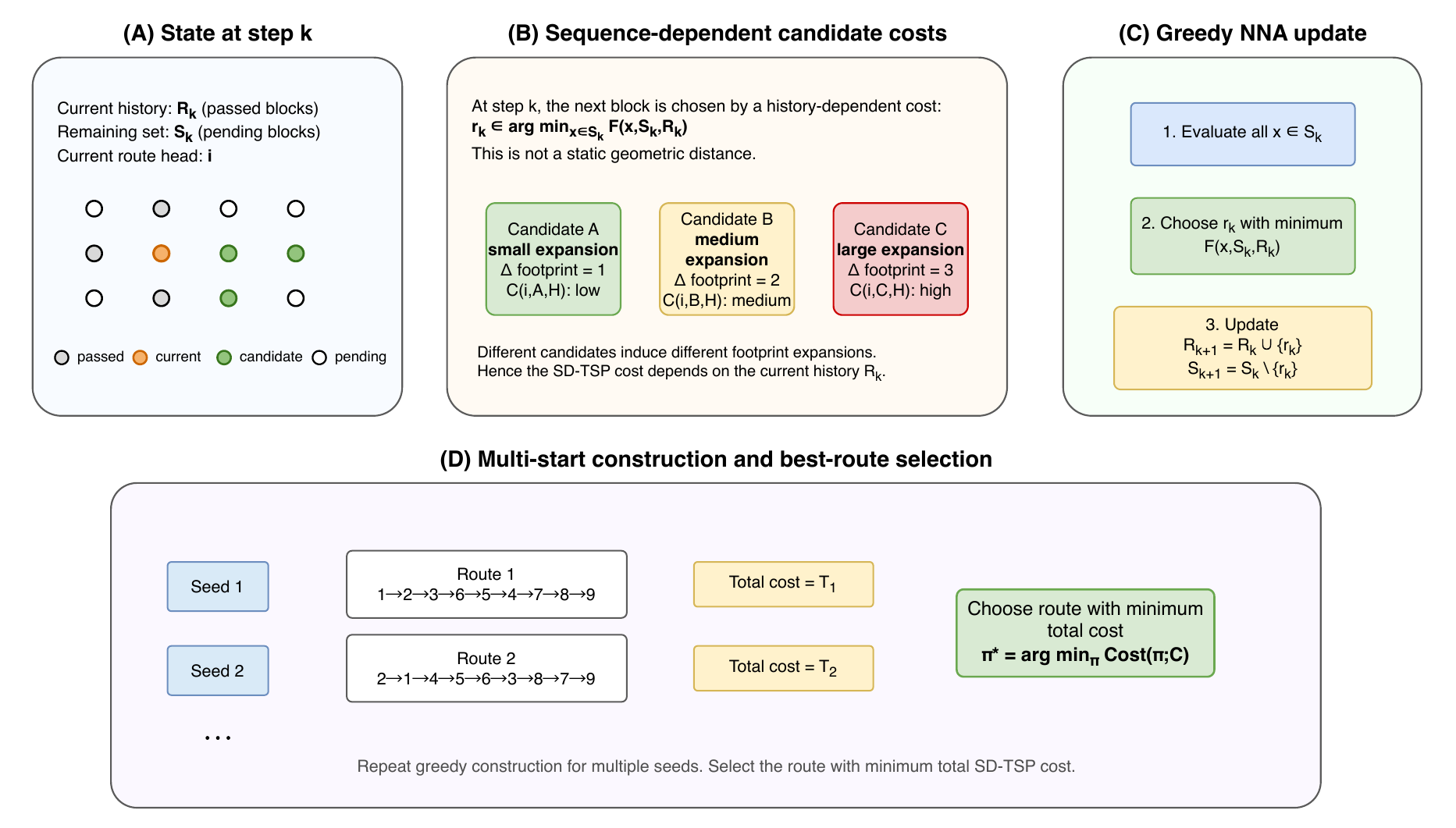}
    
    \caption{
    \textbf{History-dependent NNA procedure for solving the SD-TSP ordering problem.}
    (A) At step \(k\), the current partial route is represented by the visited-block history \(R_k\), the remaining candidate set by \(S_k\), and the current route head by the most recently selected block. This panel defines the state variables on which the next-step decision depends. 
    (B) The candidate cost is sequence dependent: for each \(x\in S_k\), the NNA evaluates a history-dependent incremental cost \(F(x,S_k,R_k)\), schematically reflecting how much the reduced-objective footprint would expand if \(x\) were selected next. The three examples illustrate that different candidates can induce different footprint expansions and therefore different SD-TSP costs, even when the geometric separation is similar. 
    (C) The greedy NNA update chooses the next block \(r_k\in\arg\min_{x\in S_k}F(x,S_k,R_k)\), then updates the visited set and remaining set according to \(R_{k+1}=R_k\cup\{r_k\}\) and \(S_{k+1}=S_k\setminus\{r_k\}\). 
    (D) To reduce seed dependence, the greedy construction is repeated from multiple initial seeds, producing multiple candidate routes; the final traversal order is the route \(\pi^*\) with minimum total sequence-dependent cost \(Cost(\pi;C)\). 
    The key distinction from a classical TSP is that the effective transition cost is not a fixed pairwise distance but depends on the evolving history of visited blocks through the reduced-objective footprint.
    }
    \label{fig:nna-flowchart}
\end{figure*}

\noindent The sequence-dependent Traveling Salesman Problem  is the combinatorial optimization problem:
\begin{equation}
    \min_\pi Cost(\pi; C),
\end{equation}
Figure~\ref{fig:nna-flowchart} provides a schematic operational interpretation of this SD-TSP formulation in the present calibration setting. Rather than assigning a fixed pairwise distance between two blocks, the figure emphasizes that the effective transition cost depends on the current traversal history through the evolving reduced-objective footprint, which is the key reason that the ordering problem is sequence dependent.

Note that when $C(i, j, H)$ depends only on $(i, j)$ (that is, $C(i, j, H) = c(i, j)$ for all histories $H$), the problem reduces to the classical Traveling Salesman Problem(TSP).

In our problem, we usually have another graph $G_p = (V_p, E_p)$, where $V_p=\{1,2,3,\dots,N\}$ is the set of qubits and  $E_p = \{(i, j) | i, j \in V_p, i \neq j\}$ is the set of two-qubit gate in the chip. 
To simplify the process of determining how to divide parameters into different blocks initially, 
we partition the parameters into $N$ blocks. 
Each block consists of one single-qubit gate parameter and between zero and four two-qubit gate parameters. 
The number of two-qubit gate parameters in one block depends on whether another single-qubit gate parameter has already been divided. 
In this case, the position of single-qubit gate parameters and the order in which all single-qubit gate parameters are traversed determine the block set $\mathcal{B}=\{B_j\}_{j=1}^M$. Therefore, the set of qubits and the vertex set in the optimization problem are identical($V_p=V$), while the edge set is typically different($E_p\neq E$).
This is because the graph chosen for the TSP problem is complete, with directed edges representing the sequence of parameter block optimizations, allowing edges to exist between any two points. 
In contrast, the graph is often local due to the positional constraints of the two-qubit gates.

In this TSP formulation, the "travel cost" between two specific blocks is not fixed but depends on the sequence of blocks that the traveler (or qubit) has previously visited. As a result, certain greedy algorithms traditionally used for solving TSP cannot be directly applied. However, the Nearest Neighbor Algorithm (NNA) emerges as a viable alternative in this context \cite{Rosenkrantz1974,Johnson1990,Pedro2013,Hansen2018}.
We do not claim that NNA is the best possible solver for a general SD-TSP. In the present setting, however, the ordering cost depends on the traversal history through the evolving reduced-objective footprint, so classical TSP heuristics that rely on fixed pairwise edge weights or precomputed distance matrices do not apply directly. More general metaheuristics---simulated annealing, genetic algorithms---can in principle be adapted, but at the expense of repeated full-route evaluations and additional tuning. NNA requires only local, incremental cost comparisons, naturally handles the history-dependent cost \(C(i,j,H)\), and is lightweight enough to embed directly in the BCD loop. We therefore adopt it here as a practical solver and leave stronger SD-TSP-specific heuristics as a direction for future work.

\begin{algorithm}[H]
  \caption{Nearest-neighbor algorithm}
  \label{alg:nna}
  \begin{algorithmic}[1]
    \Require $S$ point set; $F$ cost function
    \Ensure  $R$ route
    \State $k \gets 0$
    \State $R_k \gets \varnothing,\; S_k \gets S$
    \Repeat
      \State choose $r_k \in \operatorname*{arg\,min}\limits_{x\in S_k} F(x,S_k,R_k)$
      \State $R_{k+1} \gets R_k \cup \{r_k\},\; S_{k+1} \gets S_k \setminus \{r_k\}$
      \State $k \gets k+1$
    \Until{$S_k=\varnothing$}
  \end{algorithmic}
\end{algorithm}

Using the NNA (Alg.~\ref{alg:nna}) approach, the optimizer starts from one seed block and incrementally constructs a traversal route for the block-coordinate updates. At step \(k\), the current optimization state is described by the visited-block history \(R_k\), the remaining candidate set \(S_k\), and the current route head, as illustrated in Fig.~\ref{fig:nna-flowchart}(A). For each candidate next block \(x\in S_k\), the algorithm evaluates a history-dependent incremental cost \(F(x,S_k,R_k)\), which serves as the SD-TSP selection criterion. In the present calibration setting, this cost is not a static geometric distance; rather, it reflects how selecting \(x\) next would expand the reduced-objective footprint and thereby increase the effective per-step evaluation cost, as illustrated schematically in Fig.~\ref{fig:nna-flowchart}(B). The next block is then chosen greedily as the minimizer of this incremental cost, followed by the updates \(R_{k+1}=R_k\cup\{r_k\}\) and \(S_{k+1}=S_k\setminus\{r_k\}\), as shown in Fig.~\ref{fig:nna-flowchart}(C). Repeating this process yields one complete candidate route.

To reduce sensitivity to the initial seed, we repeat this greedy construction from multiple starting blocks and then select the route with the minimum total sequence-dependent cost, as shown in Fig.~\ref{fig:nna-flowchart}(D). In this way, the final block-update order is not determined by a single arbitrary initialization, but by a lightweight multi-start procedure that remains polynomial-time while better exploiting the topology-dependent structure of the reduced objectives. Although NNA is inherently greedy, this construction provides an efficient and implementation-friendly method for obtaining a low-cost traversal order for the BCD framework.

Here, we aim to examine the impact of the NNA algorithm on the BCD algorithm with numerical simulation. 
First, we considered the differences arising from various block strategies within the BCD algorithm. 
These differences primarily manifest in two aspects: optimization results and optimization complexity.
We randomly select block strategies from the permissible strategy space and compare them with those chosen by the NNA algorithm.
As shown in Fig.~\ref{fig:nna-bcd}(a,b), the NNA-based ordering does not degrade the final optimization quality relative to randomly selected block orders; the distribution of final-error differences remains centered close to zero and is slightly shifted in favor of BCD-NNA. At the same time, the corresponding empirical algorithmic cost ratio is consistently larger than one, showing that BCD-NNA reduces the per-epoch cost relative to random ordering. In this sense, the main benefit of the NNA ordering is to improve efficiency while retaining, and on average slightly improving, the final calibration quality.

To further benchmark the proposed framework against a broader heuristic baseline, we additionally compare BCD-NNA with a genetic algorithm (GA), as shown in Fig.~\ref{fig:nna-bcd}(c). In this comparison, the distribution of the final-error difference \( \mathrm{BCD\mbox{-}NNA} - \mathrm{GA} \) is also centered close to zero, but with a slightly negative sample mean. This indicates that BCD-NNA achieves optimization quality comparable to the GA baseline and is slightly better on average on the tested simulator instances. Since the GA does not explicitly exploit the block-local structure or the SD-TSP interpretation of the ordering problem, this comparison supports the view that the effectiveness of the proposed method comes from combining BCD with a topology-aware ordering mechanism tailored to the locality structure of the calibration task.

Taken together, Fig.~\ref{fig:nna-bcd}(a)--(c) suggests that the main advantage of the proposed method lies not in using a generic heuristic search alone, but in exploiting the locality structure of the calibration problem through the combination of block-coordinate decomposition and topology-aware ordering.
The contribution of this work is therefore not the introduction of another standalone heuristic, but the observation that casting frequency allocation in a BCD language with reduced local objectives makes the block-ordering degree of freedom explicit and amenable to topology-aware optimization. The numerical evidence shows that, within this structured formulation, the BCD-NNA combination reproducibly reduces algorithmic cost while maintaining optimization quality comparable to broader baselines.

Given the improvements of the NNA algorithm over the BCD algorithm, 
we are increasingly interested in how the optimization algorithm performs as quantum chips incorporate more qubits, 
and whether the complexity reduction achieved by the NNA algorithm changes as the number of qubits increases. 
As shown in Fig.~\ref{fig:scaling}, the optimization algorithm continues to effectively reduce the error rate of quantum circuits as the number of qubits in the system increases. 
Simultaneously, the impact of the NNA algorithm on computational complexity becomes more pronounced as the number of qubits grows.

The algorithm described above requires only information about the presence of crosstalk in the system, while the specific strength of the crosstalk is not a necessary input. Detecting the strength of crosstalk becomes relevant only under particular circumstances. Specifically, if the system exhibits neither exceptionally strong nor unexpectedly weak crosstalk, measuring the crosstalk strength does not yield significant improvements in the BCD algorithm.

However, if certain crosstalk interactions are found to be unusually strong, additional experimental methods can be used to mitigate these effects. Simultaneously, it becomes beneficial to group the parameters influenced by these strong crosstalk interactions into fewer blocks. By confining the influence of crosstalk within the parameters of a single block, the crosstalk between blocks is effectively reduced. Nevertheless, this approach increases the size of each block ($|B_j|$) and the associated coupling terms ($|C_j|$), thereby leading to a rise in the overall computational complexity.

Conversely, if certain crosstalk interactions are found to be exceptionally weak, the inclusion range of the reduced objective function 
$G_{B_j}$
can be narrowed, which decreases 
$|C_j|$ and subsequently reduces the computational complexity of the BCD algorithm. In cases where a large number of weak crosstalk interactions are present, such that the qubit system can be divided into two or more independent subsystems with negligible mutual influence, the original N-qubit system can be partitioned into smaller subsystems. Given the exponential increase in the complexity of Eq.~(\ref{eq:opt}) with the number of qubits, this partitioning significantly reduces computational complexity. Furthermore, the smaller subsystems can be processed in parallel using the BCD algorithm, further reducing the computational burden.

Adaptive treatment of crosstalk intensity can accordingly enter the framework at two levels. At the block-design level, unusually strong couplings may motivate merging more parameters into the same optimization block, whereas unusually weak couplings may justify finer decomposition. At the reduced-objective level, the assumed crosstalk graph determines which surrounding terms must be retained when evaluating \(G_{B_j}\). Thus, even if the optimization blocks remain qubit-centered, the actual local objective can still adapt to the inferred crosstalk range and strength. This is why the present framework is not intrinsically restricted to a nearest-neighbor square lattice. More generally, once the parameter locations and an assumed crosstalk interaction graph---or an effective interaction range extracted from preliminary characterization---are available, the same construction can be applied to quantum chips with different architectures and coupling layouts.

In practice, the assumed crosstalk model may be imperfect. In that case the algorithm should be interpreted as operating with an approximate reduced objective rather than an exact one. As our nonlocal-crosstalk robustness experiments indicate, moderate mismatch between the assumed and actual crosstalk structure does not necessarily invalidate the method; rather, it degrades the optimization quality gradually. This tolerance is important for practical use on heterogeneous devices, where the interaction graph may not be known exactly in advance. We therefore regard adaptive block construction and adaptive reduced-objective construction based on experimentally inferred crosstalk strength as natural extensions of the present framework, and as promising directions for future work.

\section{Conclusion}
In both near-term superconducting processors and future large-scale architectures, classical optimization remains essential for benchmarking, calibration, and retuning of device parameters.
Classical computers play a critical role in evaluating the performance of quantum processors by quantifying the gap between the realized quantum circuit and the idealized circuit implementation. 
In this work, we have established the mathematical equivalence between the Snake optimizer and Block Coordinate Descent (BCD), providing a rigorous theoretical foundation for this widely-used strategy for qubit frequency allocation. This formalization enables the application of classical optimization theory to analyze convergence properties and complexity scaling. Building on this foundation, we formulate the block ordering problem as a Sequence-Dependent Traveling Salesman Problem (SD-TSP) and solve it efficiently using the Nearest Neighbor Algorithm (NNA), achieving systematic complexity reduction compared to the graph-based heuristics (BFS, DFS) employed in the original Snake implementation.

The proposed method achieves a computational complexity of O(N) under the assumption of local crosstalk, making it highly efficient for practical scenarios. Furthermore, the approach demonstrates strong robustness against random noise, ensuring reliable performance in realistic, noisy quantum environments. However, the method's performance can deteriorate in the presence of strong nonlocal crosstalk interactions, which underscores the limitations of the current algorithm and the need for further refinements.
The paper progresses from a mathematical reformulation that identifies frequency allocation as a BCD problem with reduced local surrogates, through a topology-aware ordering strategy that exploits the resulting block structure, to numerical validation under a physics-motivated surrogate model. Each stage builds on the preceding one: the reformulation makes convergence, block-size trade-offs, and ordering costs precisely discussable; the NNA ordering is a concrete improvement visible only within that reformulation; and the simulations confirm that the structural improvement translates into measurable practical gains.

From an FTQC-oriented perspective, the relevance of the proposed framework does not rely on the NISQ label itself, but rather on three structural conditions: sparse device connectivity, predominantly local crosstalk, and the availability of block-local surrogate objectives that can be evaluated more cheaply than the full-chip objective. Whenever these conditions remain valid, the BCD-NNA strategy can still be useful beyond the NISQ regime---for example, in physical-qubit frequency tuning prior to logical operation, in local retuning of modules or patches within larger architectures, and in repeated recalibration under device drift. The main value of the method is therefore as a scalable, locality-aware calibration primitive that is not tied to any particular processor size.

At the same time, fully fault-tolerant architectures will likely require extensions of the present formulation. In large-scale FTQC systems, the most relevant performance target may no longer be a circuit-level physical error metric alone, but rather logical-level quantities such as code-cycle performance, syndrome-informed diagnostics, or logical error budgets subject to global architectural constraints. In such settings, the block definition, surrogate objective, and ordering cost may all need to be redesigned to reflect logical-layer structure and cross-module couplings. We therefore view the present work as a foundation for a broader class of topology-aware calibration and retuning algorithms that could be adapted to future fault-tolerant quantum computing stacks.

To address these challenges, future research will focus on the development of advanced optimization techniques. Specifically, adaptive crosstalk detection mechanisms will be explored to dynamically identify and mitigate the effects of nonlocal crosstalk. Additionally, parallel optimization strategies will be incorporated to enhance the scalability of the algorithm, enabling it to handle larger quantum systems more effectively. These advancements aim to bridge the gap between theoretical models and practical implementations, further improving the calibration and benchmarking of quantum processors in both near-term and scalable fault-tolerant settings.

\begin{acknowledgements}
This work is supported by Shanghai Municipal Science and Technology (Grant No. 25LZ2600200), National Natural Science Foundation of China (Grants No. 92365111), Beijing Natural Science Foundation (Grants No.~Z220002), and the Innovation Program for Quantum Science and Technology (Grant No.~2021ZD0302400).
\end{acknowledgements}

\appendix

\section{Error simulator and numerical simulation details} \label{app:simulator}

\subsection{General structure of the simulator} 

The numerical simulations employ a physics-motivated error simulator for qubit frequency allocation in superconducting quantum processors. To avoid overemphasizing implementation details, we describe the simulator at the level of effective error terms rather than code-level realization. The simulator first produces a circuit-level error estimator
\begin{equation}
    \mathcal{E}_{\mathrm{circ}}(f)
    =
    \sum_{i\in\mathcal{Q}} \mathcal{E}_i^{\mathrm{single}}(f)
    +
    \sum_{\langle i,j\rangle\in\mathcal{E}} \mathcal{E}_{ij}^{\mathrm{near}}(f)
    +
    \sum_{(i,j)\in\mathcal{P}_{\mathrm{far}}} \mathcal{E}_{ij}^{\mathrm{far}}(f).
    \label{eq:app-total}
\end{equation}
Here the first sum contains single-body terms, the second sum contains nearest-neighbor multi-body terms determined by the chip topology, and the third sum contains non-nearest-neighbor corrections that are switched on only in the model-mismatch tests. This form makes explicit which parts of the objective are local and topology-aware, and which parts correspond to neglected long-range effects.

\subsection{Single-body terms} 

The single-body contribution for qubit \(i\) is decomposed as
\begin{equation}
    \mathcal{E}_i^{\mathrm{single}}(f)
    =
    \mathcal{E}_i^{(\phi)}(f_i)
    +
    \mathcal{E}_i^{(T_1)}(f_i).
\end{equation}
The term \(\mathcal{E}_i^{(\phi)}\) describes one-qubit dephasing-related error. Its frequency dependence reflects the fact that flux-tunable superconducting qubits become more sensitive to flux noise in some operating regions than in others. The term \(\mathcal{E}_i^{(T_1)}\) describes one-qubit relaxation-related error and is modeled through a frequency-dependent relaxation profile, which phenomenologically captures the variation of energy relaxation with qubit frequency.

In the simulator, these terms are smooth but nontrivial functions of frequency and are sampled with qubit-dependent coefficients from fixed intervals. Therefore, different random simulator instances correspond to different effective devices with different dephasing and relaxation landscapes.

\subsection{Nearest-neighbor multi-body terms} 

The nearest-neighbor contribution is written schematically as
\begin{equation}
    \mathcal{E}_{ij}^{\mathrm{near}}(f)
    =
    \mathcal{E}_{ij}^{(\mathrm{coll})}(f_i,f_j)
    +
    \mathcal{E}_{ij}^{(\mathrm{2q},\phi)}(f)
    +
    \mathcal{E}_{ij}^{(\mathrm{2q},T_1)}(f)
    +
    \mathcal{E}_{ij}^{(\mathrm{fluxXT})}(f).
\end{equation}

The term \(\mathcal{E}_{ij}^{(\mathrm{coll})}\) models frequency-collision or near-collision errors between coupled computational elements. In physical terms, if two nearby transitions become too close in frequency, unwanted hybridization or parasitic interaction can increase the effective gate error. In the simulator this is represented by a sharply enhanced penalty when relevant frequency differences become small. Thus, frequency collisions are not introduced as separate random events, but as explicit deterministic contributions to the underlying noiseless simulator.

The terms \(\mathcal{E}_{ij}^{(\mathrm{2q},\phi)}\) and \(\mathcal{E}_{ij}^{(\mathrm{2q},T_1)}\) describe the dephasing and relaxation accumulated during a two-qubit operation. These terms depend on both the idle frequencies and the effective interaction frequency because the two-qubit gate follows a frequency trajectory in time. In the simulator, they are represented by effective trajectory-dependent penalties that increase when the gate passes through unfavorable spectral regions.

The term \(\mathcal{E}_{ij}^{(\mathrm{fluxXT})}\) models flux crosstalk. In the baseline setting of the numerical simulations, this contribution is assumed to be predominantly local, namely dominated by nearest-neighbor couplings on the chip graph. This choice is consistent with the intended use of the reduced local objective in the BCD framework.

\subsection{Nonlocal multi-body correction terms} 

To test the robustness of the method against model mismatch, we further consider a nonlocal contribution
\begin{equation}
    \mathcal{E}_{ij}^{\mathrm{far}}(f),
    \qquad (i,j)\in\mathcal{P}_{\mathrm{far}},
\end{equation}
where \(\mathcal{P}_{\mathrm{far}}\) contains qubit pairs or gate pairs that are not nearest neighbors in the chip topology. This term is used only in the robustness experiments associated with Fig.~\ref{fig:nonlocal-crosstalk}. Its role is not to redefine the baseline model, but to quantify the performance loss that occurs when the optimizer assumes a local crosstalk model while the underlying device contains additional long-range interactions.

In particular, our discussion of flux crosstalk separates two levels: the baseline simulator assumes that flux crosstalk is local and short-ranged, while the robustness study introduces nonlocal flux-crosstalk-like corrections and compares optimization with and without explicitly accounting for them. This is the sense in which we test whether the crosstalk range is fully captured by the reduced optimization model.

\subsection{Noisy objective evaluations and stochastic measurement noise} 

The stochastic noise used during optimization is intended to mimic uncertainty from finite experimental measurements. It is therefore introduced at the level of objective-function evaluation, not by directly modifying the underlying physical device instance. If \(\mathcal{E}_{\mathrm{circ}}(f)\) is the noiseless simulator output, then the optimizer accesses
\begin{equation}
    \tilde{\mathcal{E}}_{\mathrm{circ}}(f)=\mathcal{E}_{\mathrm{circ}}(f)\bigl(1+\xi\bigr)+\delta_{\mathrm{miss}}(f),
\end{equation}
where \(\xi\) is a random variable describing measurement-like fluctuations and \(\delta_{\mathrm{miss}}(f)\) is the deterministic model-mismatch contribution arising when nonlocal crosstalk is present in the simulator but neglected by the optimizer.

Thus, the random noise is added to the circuit-level objective value returned to the optimizer at each query. This corresponds to the practical situation in which the optimization algorithm receives an imperfect estimate of circuit quality from a finite measurement process. By contrast, the final error values reported in the figures are evaluated from the underlying noiseless simulator \(\mathcal{E}_{\mathrm{circ}}(f)\), so that the comparison reflects optimization quality rather than one-shot readout fluctuations.

\subsection{Meaning of random instances, random initial points, and plotted distributions} 

The numerical experiments involve three distinct sources of randomness:
\begin{enumerate}
    \item \textbf{Random simulator instances:} the coefficients and characteristic parameters of the physical error terms are sampled to generate different device instances.
    \item \textbf{Random initial points:} for a fixed simulator instance, the optimization starts from different initial frequency allocations within the feasible domain.
    \item \textbf{Random measurement noise:} each objective-function query during the optimization can be perturbed by a stochastic multiplicative fluctuation.
\end{enumerate}

Because the distributions shown in Fig.~\ref{fig:optimization-performance}, Fig.~\ref{fig:nonlocal-crosstalk}, and Fig.~\ref{fig:nna-bcd} combine nonconvex optimization, device-to-device variation, and noisy objective queries, they should not be interpreted as simple repeated measurements around one fixed mean value. Accordingly, a Gaussian distribution is not expected in general. The plotted density functions summarize the empirical distribution of optimization outcomes under this full simulation protocol.

\subsection{Definition of the reported average gate error rate} 

For self-contained reference, we restate the conversion from the simulator's circuit-level output to the reported gate-level metric.
The quantity directly produced by the simulator and used during optimization is the circuit-level error estimator \(\mathcal{E}_{\mathrm{circ}}(f)\), obtained by summing the relevant single-body, nearest-neighbor, and nonlocal contributions. We interpret the corresponding circuit success probability as
\begin{equation}
    P_{\mathrm{circ}}(f)=1-\mathcal{E}_{\mathrm{circ}}(f).
\end{equation}

To report results on a gate-level scale, we convert this circuit-level quantity into an effective average gate error rate. Suppose that the circuit template used in the numerical evaluation contains \(G\) effective gate operations and that these gates are assigned a common average success probability \(\bar{p}_g(f)\). Then we define
\begin{equation}
    P_{\mathrm{circ}}(f)=\bar{p}_g(f)^G,
\end{equation}
which yields
\begin{equation}
    \bar{p}_g(f)=\bigl(1-\mathcal{E}_{\mathrm{circ}}(f)\bigr)^{1/G}.
\end{equation}
Therefore, the reported average gate error rate is
\begin{equation}
    \bar{\varepsilon}_g(f)=1-\bar{p}_g(f)
    =1-\bigl(1-\mathcal{E}_{\mathrm{circ}}(f)\bigr)^{1/G}.
    \label{eq:app-avg-gate}
\end{equation}

This is the quantity shown in the main-text figures. In the small-error regime one has the approximation \(\bar{\varepsilon}_g(f)\approx \mathcal{E}_{\mathrm{circ}}(f)/G\), but the numerical results reported in the manuscript are obtained using the product-form definition in Eq.~(\ref{eq:app-avg-gate}), not this linearized approximation.

\subsection{Definition of the complexity metrics used in the numerical figures} 

For the numerical comparisons of traversal strategies, we distinguish between two complexity notions. The first is an empirical algorithmic cost proxy used to characterize the implemented block-wise optimizer. For a block \(B_i\), this proxy combines the local search cost of the inner derivative-free optimizer, scaling as \(3^{|B_i|}\) per local step, with the effective cost of evaluating the reduced local objective. The latter is represented by the number of retained simulator terms in that reduced objective, denoted \(N_{\mathrm{term}}(B_i)\). For fixed local iteration budget \(S\), this gives
\begin{equation}
    \mathcal{C}^{\mathrm{emp}}_i = S\,3^{|B_i|}N_{\mathrm{term}}(B_i).
\end{equation}

The second is the search-space complexity model used for scaling analysis, in which the block search cost is represented by \(k^{|B_i|}\) and the reduced-objective evaluation cost is represented schematically by a footprint-dependent factor such as \(t^{|C_i|}\). This yields a total cost of the form
\begin{equation}
    \mathcal{T}_{\mathrm{search}}
    =
    \sum_i S\,k^{|B_i|}t^{|C_i|}.
\end{equation}

The complexity ratios shown in the manuscript are therefore algorithmic cost ratios, not hardware-specific wall-clock timing ratios. This choice allows the comparison of ordering strategies to remain independent of software environment, processor load, and implementation details not central to the method itself.

\bibliography{ref.bib}

\end{document}